\documentclass[preprint]{iucr}              

\usepackage{amsmath, amsthm, amssymb, amsfonts, textcomp, gensymb, subfigure, import, color, siunitx, float, bm, graphicx, url, comment}

     \journalcode{J}              

\begin{document}                  



\title{Refinements for Bragg coherent X-ray diffraction imaging: Electron backscatter diffraction alignment and strain field computation}


\author[a]{David}{Yang}
\author[a]{Mark}{T. Lapington}
\author[a]{Guanze}{He}
\author[a]{Kay}{Song}
\author[b]{Minyi}{Zhang}
\author[b]{Clara}{Barker}
\author[d]{Ross}{J. Harder}
\author[d]{Wonsuk}{Cha}
\author[d]{Wenjun}{Liu}
\author[c]{Nicholas}{W. Phillips}
\cauthor[a]{Felix}{Hofmann}{david.yang@eng.ox.ac.uk  \linebreak E-mail: felix.hofmann@eng.ox.ac.uk}

\aff[a]{Department of Engineering Science, University of Oxford, Parks Road, Oxford, OX1~3PJ, \country{UK}}
\aff[b]{Department of Materials, University of Oxford, Parks Road, Oxford, OX1~3PH, \country{UK}}
\aff[c]{Paul Scherrer Institut, 5232 Villigen PSI, \country{Switzerland}}
\aff[d]{Advanced Photon Source, Argonne National Laboratory, Lemont, IL 60439, \country{USA}}









\maketitle                        

\section{Abstract}
Bragg coherent X-ray diffraction imaging (BCDI) allows the three-dimensional (3D) measurement of lattice strain along the scattering vector for specific microcrystals. If at least three linearly independent reflections are measured, the  3D variation of the full lattice strain tensor within the microcrystal can be recovered. However, this requires knowledge of the crystal orientation, which is typically attained via estimates based on crystal geometry or synchrotron micro-beam Laue diffraction measurements. Here, we present an alternative method to determine the crystal orientation for BCDI measurements, by using electron backscatter diffraction (EBSD) to align Fe-Ni and Co-Fe alloy microcrystals on three different substrates. The orientation matrix is calculated from EBSD Euler angles and compared to the orientation determined using micro-beam Laue diffraction. The average angular mismatch between the orientation matrices is less than $\sim6$ degrees, which is reasonable for the search for Bragg reflections. We demonstrate the use of an orientation matrix derived from EBSD to align and measure five reflections for a single Fe-Ni microcrystal using multi-reflection BCDI. Using this dataset, a refined strain field computation based on the gradient of the complex exponential of the phase is developed. This approach is shown to increase accuracy, especially in the presence of dislocations. Our results demonstrate the feasibility of using EBSD to pre-align BCDI samples and the application of more efficient approaches to determine the lattice strain tensor with greater accuracy.


\section{Introduction} \label{section:introduction}
Bragg coherent X-ray diffraction imaging (BCDI) allows three-dimensional (3D) nanoscale strain measurements, with a typical spatial resolution of a few tens of nanometres and a strain resolution on the order of $\sim$2$\times10^{-4}$ \cite{Hofmann2017a}. BCDI has been applied to study crystal defects and lattice strain in a variety of materials, including noble metals \cite{Robinson2001}, alloys \cite{Kawaguchi2021}, geological compounds \cite{Yuan2019a}, semiconductors \cite{Lazarev2018}, and functional materials \cite{Dzhigaev2021}. An advantage of using BCDI is the ability to study 3D volumes up to $1\ \mathrm{\mu m}$ in size at ambient conditions. This has enabled BCDI to become an essential tool for probing how lattice strains evolve in \textit{in situ} and \textit{operando} studies, for example in battery charging \cite{Singer2018}, thermal diffusion \cite{Estandarte2018}, dissolution \cite{Clark2015} and catalytic oxidation \cite{Carnis2021b}.

BCDI involves fully illuminating a crystalline sample with a coherent X-ray beam, and positioning the diffractometer such that the Bragg condition is met for a specific $hkl$ reflection. The outgoing wave vector produces a diffraction pattern that is collected on a pixelated area detector positioned in the far-field (Fraunhofer regime). By rotating the sample through the Bragg condition, a 3D coherent X-ray diffraction pattern (CXDP) is recorded as different parts of the 3D Bragg peak sequentially intersect the Ewald sphere in reciprocal space, which is projected onto the detector. If the CXDP is oversampled by at least twice the Nyquist frequency \cite{Sayre1952}, iterative phase retrieval algorithms can be used to recover the phase \cite{Fienup1982}. The amplitude and phase in reciprocal space are related to the real-space object via an inverse Fourier transform \cite{Miao2000b} followed by a space transformation from detector conjugated space to orthogonal laboratory or sample space \cite{Yang2019,Maddali2020,Li2019b}. The real space amplitude, $\mathbf{\rho(r)}$, where $\mathbf{r}$ is the position vector, is proportional to the effective electron density of the crystalline volume associated with the particular crystal reflection. The real space phase, $\mathbf{\psi(r)}$, corresponds to the projection of the lattice displacement field, $\mathbf{u(r)}$, onto the Bragg vector, $\mathbf{Q_\mathit{hkl}}$, of a specific $hkl$ crystal reflection:

\begin{equation} \label{eq:phase}
    \mathbf{\psi_\mathit{hkl}(r)} = \mathbf{Q_\mathit{hkl}}\cdot\mathbf{u(r)}
\end{equation}

Since the development of BCDI in the early 2000s, most experiments feature the measurement of a single reflection, providing only one component of the strain tensor. However, the analysis of a single strain component can be ambiguous as different information is obtained for different reflections \cite{Yang2021}. If at least three linearly independent reflections are measured, the full 3D strain tensor can be calculated. Before 2017, only three experiments \cite{Beitra2010,Newton2010,Ulvestad2015b} reported measuring more than one reflection on a single crystal. This is not surprising as multi-reflection BCDI (MBCDI) experiments require prior knowledge of the crystal orientation \cite{Newton2010}, or scanning extensive volumes of reciprocal space until two reflections are found, upon which further reflections can then be located.

The development of a micro-beam Laue X-ray diffraction pre-alignment procedure in 2017 \cite{Hofmann2017b} enabled the direct determination of the crystal orientation matrix, such that crystals could be reliably pre-aligned for MBCDI. Recently, a double-bounce Si(111) monochromator that allows Laue X-ray diffraction to be performed has been commissioned at BCDI beamline 34-ID-C at the Advanced Photon Source (APS), Argonne National Laboratory, USA \cite{Pateras2020a}. Another method to determine the orientation of a sample is by indexing pole figures \cite{Richard2018}, however, this method requires a Bragg peak with known Miller indices to be found. The indexing is performed using texture analysis and relies on the samples to be well-faceted to produce truncation rods in reciprocal space that are perpendicular to facet surfaces. These pre-alignment protocols have not only led to the increased popularity of MBCDI for determination of the full strain tensor with respect to an arbitrary reference \cite{Yang2022,Hofmann2020,Hofmann2017a,Hofmann2018,Phillips2020}, but also enabled simultaneous, multi-Bragg peak phase retrieval procedures to increase reconstruction quality \cite{Newton2020,Gao2021,Wilkin2021}.

Here we present an alternative method of pre-determining crystal orientation for MBCDI alignment without relying on synchrotron X-rays. We use electron backscatter diffraction (EBSD) to determine the orientation \cite{Adams1993} of randomly-oriented Fe-Ni and Co-Fe microcrystals on three different sapphire substrates. EBSD instruments are much more widespread and accessible than synchrotron instruments, and can be used as a valuable pre-screening tool for BCDI. EBSD measurements can produce 2D orientation maps with a high spatial resolution of $\sim10\ \mathrm{nm}$, thus enabling the selection of specific crystals with particular orientations or features such as twin domains. This allows the user to preserve synchrotron beamtime for BCDI measurements rather than performing pre-orientation measurements and analysis at the beamline. We compare orientation matrices found by EBSD to those measured by micro-beam Laue diffraction and the ultimately measured reflection positions in MBCDI. Using the pre-determined EBSD orientation matrix, we measured five crystal reflections for a Fe-Ni microcrystal (Fig. \ref{fig:Tensor}) and determined its full strain and rotation tensors with respect to the average lattice of the crystal. We also implement an alternative approach, by using use the complex component of the phase, rather than the phase alone, to efficiently calculate the phase derivatives required for the strain tensor determination and more accurately interpolate the recovered phase to sample coordinates.


\section{Experimental methodology} \label{section:experimental_methodology}
\subsection{Microcrystal fabrication} \label{section:fabrication}
\begin{figure} 
    \centering
    \includegraphics[height=15cm]{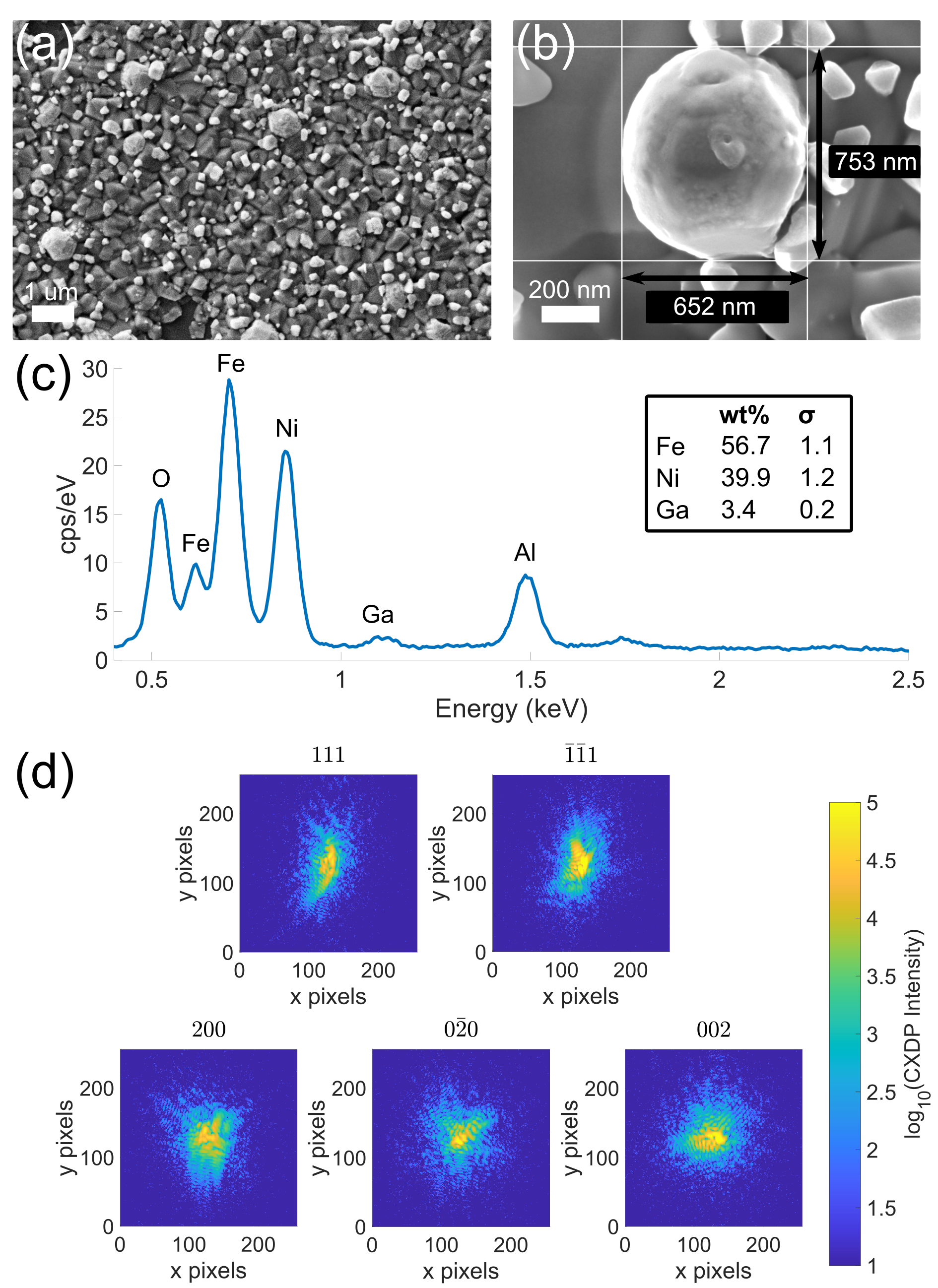}
    \caption{(a) Dewetted Fe-Ni alloy microcrystals on a sapphire substrate. (b) Fe-Ni microcrystal 1B is used for the computation of strain and rotation tensors. (c) EDX spectrum for crystal 2B on the substrate, which is similar across all crystals on the substrate. Only the L-lines for the most pronounced elements in the crystal are shown above. The composition excludes the Al and O substrate peaks. The Ga impurity is due to FIB milling around the crystal vicinity \cite{Hofmann2017a}. (d) Central slices of the CXDPs for each reflection, measured for crystal 1B.}
    \label{fig:SEM_pic}
\end{figure}

Samples were produced by sputter deposition of a thin film onto a single crystal sapphire wafer (C-plane orientation). One substrate with a film thickness of 375 nm was produced for the Fe-Ni microcrystals. It was dewetted in a vacuum furnace purged with a 5\% hydrogen, balance Argon, gas mixture at 1250\degree C for 24 hours. The resulting crystals exhibit a faced-centred cubic (fcc) structure, range from 0.5 to 1.5 $\mathrm{\mu m}$ in size, (Fig. \ref{fig:SEM_pic}(a)) and adhere to the substrate surface. The substrate was cleaved to make substrates 1 and 2, both containing Fe-Ni microcrystals. Substrate 3 contained Co-Fe microcrystals that were produced in a similar way. The procedure and details for substrate 3 can be found elsewhere \cite{Yang2022}.

Each substrate was coated with 10 nm of amorphous carbon via thermal evaporation using a Leica ACE600 coater to assist with scanning electron microscope (SEM) imaging. To facilitate reliable measurement of multiple reflections from a specific microcrystal, a ZEISS NVision 40 Ga ion FIB was used to remove the surrounding crystals within a 40 $\mathrm{\mu m}$ radius using currents from 6 nA to 150 pA, and acceleration voltage of 30 kV. Only SEM imaging was used to position the FIB milling scan to prevent large lattice strains caused by FIB imaging \cite{Hofmann2017a}. The isolated crystals on each substrate are shown in Fig. \ref{fig:Angle_maps}. Crystal 1B (Fig. \ref{fig:SEM_pic}(b)) was used for the computation of the strain and rotation tensors (Fig. \ref{fig:Tensor}).

Energy-dispersive X-ray spectroscopy (EDX) was used to determine the elemental composition of each crystal (Fig. \ref{fig:SEM_pic}(c)). EDX showed homogeneous distribution of all elements throughout the dewetted crystals (see Appendix \ref{appendix:EDX_results}). EDX was performed on a ZEISS Merlin using an Xmax 150 detector (Oxford Instruments) using an elliptical region encapsulating crystal 2B on the substrate for 16 seconds with an accelerating voltage of 10 kV.

\subsection{Electron backscatter diffraction} \label{section:EBSD}
\begin{figure} 
    \centering
    \includegraphics[height=9.5cm]{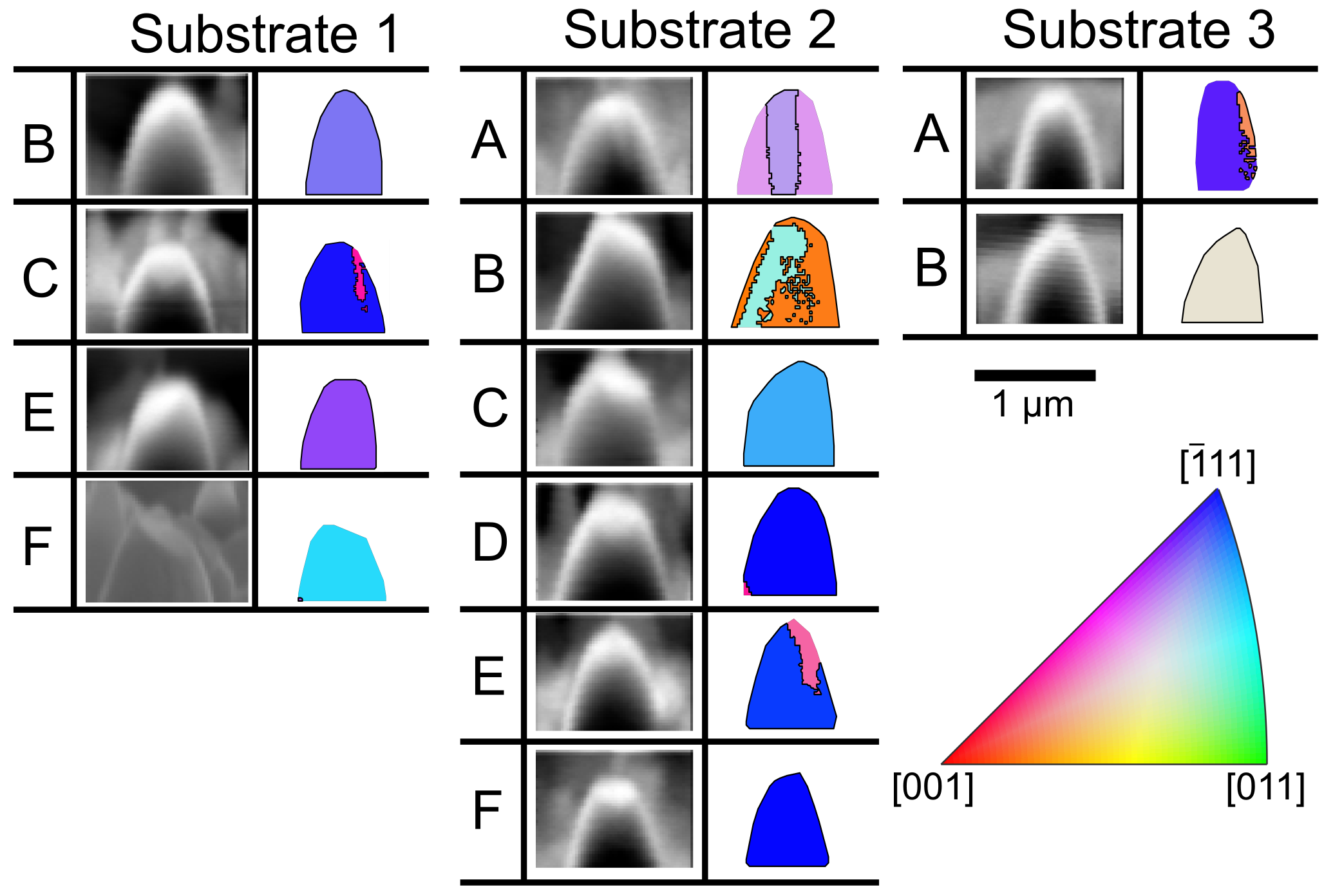}
    \caption{SEM image and inverse pole figure (IPF) maps for each crystal from EBSD. The colour-coding shows the out-of-plane crystal orientation. The maps allow for different grains to be identified. For each crystal, the region surrounded by thick black lines corresponds to the grain orientation used for the computation of the orientation matrix. For some samples (i.e. 1F, 3A) only a small region on the IPF map corresponded to the grain measured at the synchrotron. EBSD samples the orientation of a cylindrical volume with a height up to $40 \mathrm{nm}$ below the surface \cite{Dingley2004}.}
    \label{fig:Angle_maps}
\end{figure}

\begin{figure}
    \centering
    \includegraphics[height=7cm]{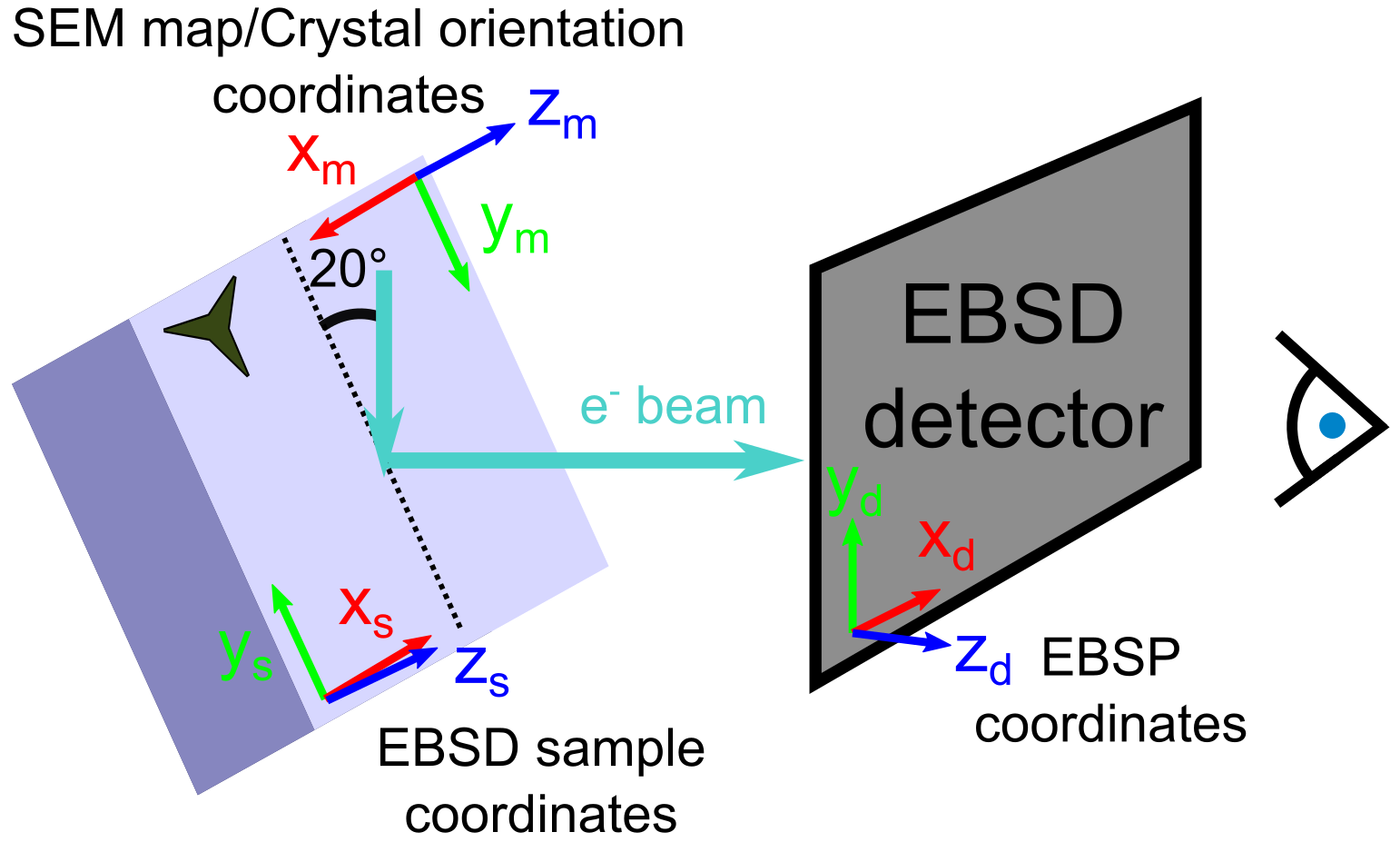}
    \caption{The position of the substrate in the laboratory frame for EBSD measurements is shown with relevant coordinate systems. The 3-pointed star represents the orientation of a sample feature on the purple substrate. The EBSD detector coordinate system ($\mathrm{x_d}$, $\mathrm{y_d}$, $\mathrm{z_d}$) describes the EBSP coordinates. The EBSD sample coordinates ($\mathrm{x_s}$, $\mathrm{y_s}$, $\mathrm{z_s}$) correspond to the EBSD scan points. The SEM map coordinates ($\mathrm{x_m}$, $\mathrm{y_m}$, $\mathrm{z_m}$) that show how EBSPs overlay on SEM maps. Here, the Euler angles output by the EBSD software are with respect to the SEM map coordinates. This follows the same convention in Britton \textit{et al.} \cite{Britton2016}.}
    \label{fig:EBSD_orientation}
\end{figure}

Crystal orientation was determined by EBSD using a ZEISS Merlin equipped with a Bruker Quantax EBSD system and a Bruker eFlash detector tilted at $4 \degree$. Electron backscatter patterns (EBSPs) were recorded with the sample tilted at $70 \degree$ (Fig. \ref{fig:EBSD_orientation}) using an accelerating voltage of 30 kV and a current of 15 nA. EBSPs were $800\times600$ pixels and a step size of 19.8 nm was used between consecutive points on the sample. The diffraction patterns were indexed and the Euler angles were extracted for each pattern using the Bruker ESPRIT 2.1 EBSD software. The Euler data was exported and analysed using MTEX, a MATLAB toolbox for texture analysis \cite{Bachmann2011}, to produce inverse pole figure (IPF) maps for all crystals, shown in Fig. \ref{fig:Angle_maps}. 

Here we use the Bunge convention \cite{Bunge1982} to describe each crystal orientation (crystal frame) relative to the substrate (sample frame). The crystal orientation matrix, $\mathbf{UB}$, is composed of $\mathbf{U}$, which describes the rotation of the crystal reference frame, and $\mathbf{B}$ (Eq. \ref{eq:B}), which characterises the unit cell parameters.

Using the same convention as Britton \textit{et al.} \cite{Britton2016}, the unit cell has lattice vectors $\mathbf{a}$, $\mathbf{b}$ and $\mathbf{c}$, each with lengths $a$, $b$, and $c$, respectively. Angle $\alpha$ describes the angle between the $\mathbf{b}$ and $\mathbf{c}$ axes, $\beta$ the angle between the $\mathbf{c}$ and $\mathbf{a}$ axes, and $\gamma$ the angle between the $\mathbf{a}$ and $\mathbf{b}$ axes. $\mathbf{B}$ is used to transform the lattice base vectors to Cartesian base vectors:

\begin{equation} \label{eq:B}
    \mathbf{B} = \begin{bmatrix}
        a\frac{f}{\mathrm{sin(\alpha)}} & 0 & 0 \\
        a\frac{\mathrm{cos(\gamma)-cos(\alpha)cos(\beta)}}{\mathrm{sin(\alpha)}} & b\mathrm{sin(\alpha)} & 0 \\
        a\mathrm{cos(\beta)} & b\mathrm{cos(\alpha)} & c\\
        \end{bmatrix}, 
\end{equation}

where

\begin{equation} \label{eq:f}
    f = \sqrt{1-\mathrm{[cos(\alpha)]^2-[cos(\beta)]^2-[cos(\gamma)]^2+2cos(\alpha)cos(\beta)cos(\gamma)}}.
\end{equation}

Since all crystals in this study have a fcc structure, $\mathbf{B}$ is the $3 \times 3$ identity matrix multiplied by the lattice constant.

$\mathbf{UB}$ provides the direction and radial position of specific $hkl$ reflections, $\mathbf{H_{\mathit{hkl}}}$, in laboratory coordinates, \cite{Busing1967}

\begin{equation} \label{eq:UB}
    \mathbf{H_{\mathit{hkl}}} = \mathbf{UB}\begin{bmatrix} h \\ k \\ l \end{bmatrix}.
\end{equation}

Several different coordinate frames are used in EBSD that refer to different aspects of the measurement (Fig. \ref{fig:EBSD_orientation}). In this paper, all coordinate systems and rotation matrices will be right-handed and the same notation as in Britton \textit{et al.} \cite{Britton2016}. For EBSD, the following subscripts describe specific coordinate systems:

\begin{itemize}
    \item `d' is the detector frame that describes the EBSPs.
    \item `s' is the EBSD sample frame that is related to the detector frame by sample. ($\theta_{\mathrm{sample}}$) and detector ($\theta_{\mathrm{detector}}$) tilts about the $\mathrm{x}$-axis. The $\mathrm{x_s}$ and $\mathrm{y_s}$ axes correspond to the directions of EBSD scan points. 
   \item `m' is the SEM map frame. This corresponds to how the EBSPs overlay on SEM maps, and thus how the EBSD orientation is referenced.
\end{itemize}

The EBSD software returns the orientation matrix from EBSD measurements in the SEM map frame (Fig. \ref{fig:EBSD_orientation}). The corresponding orientation matrix is referred to as $\mathbf{U_{EBSD,\ m}}$. It can be constructed from a series of rotations using Euler angles, where each angle describes a rotation about a coordinate axis. Here we use right-handed rotation matrices to describe vector rotations about the $\mathrm{x}$-axis (Eq. \ref{eq:R_x}),

\begin{equation} \label{eq:R_x}
    \mathbf{R\mathrm{_x(\theta)}} = \begin{bmatrix}
        1 & 0 & 0 \\
        0 & \mathrm{cos(\theta)} & -\mathrm{sin(\theta)} \\
        0 & \mathrm{sin(\theta)} & \mathrm{cos(\theta)} \\
        \end{bmatrix},
\end{equation}

and the z-axis (Eq. \ref{eq:R_z}),

\begin{equation} \label{eq:R_z}
    \mathbf{R\mathrm{_z(\theta)}} = \begin{bmatrix}
        \mathrm{cos(\theta)} & -\mathrm{sin(\theta)} & 0 \\
        \mathrm{sin(\theta)} & \mathrm{cos(\theta)} & 0 \\
        0 & 0 & 1 \\
        \end{bmatrix}. 
\end{equation}

and Bunge convention Euler angles, $\phi_1$, $\Phi$, and $\phi_2$ \cite{Britton2016}. Equivalently, Eq. \ref{eq:R_x} and Eq. \ref{eq:R_z} correspond to left-handed rotations through angle $\theta$ for the coordinate system. To transform vectors from the crystal coordinate frame to the laboratory frame, we use Eq. \ref{eq:U_EBSDm_BB} \cite{Britton2016},

\begin{equation} \label{eq:U_EBSDm_BB}
    \mathbf{U_{EBSD,\ m}}^\top = \mathbf{R\mathrm{_z(-\phi_2)}}\mathbf{R\mathrm{_x(-\Phi)}}\mathbf{R\mathrm{_z(-\phi_1)}}.
\end{equation}

We note that the Euler angles in Eq. \ref{eq:U_EBSDm_BB} are negative because the original angles as defined are for left-handed rotation matrices. First a rotation of $-\phi_1$ is applied about the original $\mathrm{z}$ axis, followed by a rotation of $-\Phi$ about the new $\mathrm{x}$ axis, and finally a rotation of $-\phi_2$ about the new $\mathrm{z}$ axis. For consistency with the convention here \cite{Busing1967}, we express Eq. \ref{eq:U_EBSDm_BB} as

\begin{equation} \label{eq:U_EBSDm}
    \mathbf{U_{EBSD,\ m}} = \mathbf{R\mathrm{_z(\phi_1)}}\mathbf{R\mathrm{_x(\Phi)}}\mathbf{R\mathrm{_z(\phi_2)}}.
\end{equation}

Here, the EBSD software already accounts for instrument tilts, $\theta_{\mathrm{sample}}$ and $\theta_{\mathrm{detector}}$, and returns Euler angles in the SEM map frame (subscript m) (Fig. \ref{fig:EBSD_orientation}). 

To input the orientation matrix into the \textit{spec} orientation calculator on beamline 34-ID-C, we must define two $hkl$ reflections corresponding to out-of-plane (Eq. \ref{eq:primary}) and in-plane reflections (Eq. \ref{eq:secondary}) \cite{Hofmann2017b}. One concern is the consistency in the indexation of crystals at the 34-ID-E Laue instrument and in EBSD measurements. Due to the cubic structure of the present crystals, there are equivalent orientation matrices that differ by $90\degree$ rotations about the crystal axes. These rotations are accounted for using a $\mathbf{R\mathrm{_{crystal}}}$ rotation matrix that captures rotations by $90(n_{x,y,z})\degree$, where $n_{x,y,z} \in \{-1,0,1,2\}$, about the $\mathrm{x}$, $\mathrm{y}$, and $\mathrm{z}$ axes (Eq. \ref{eq:R_crystal}),

\begin{equation} \label{eq:R_crystal}
    \mathbf{R\mathrm{_{crystal}}} = \mathbf{R_x}(90(n_x)\degree)\mathbf{R_y}(90(n_y)\degree)\mathbf{R_z}(90(n_z)\degree).
\end{equation}

\begin{figure}
    \centering
    \includegraphics[height=12cm]{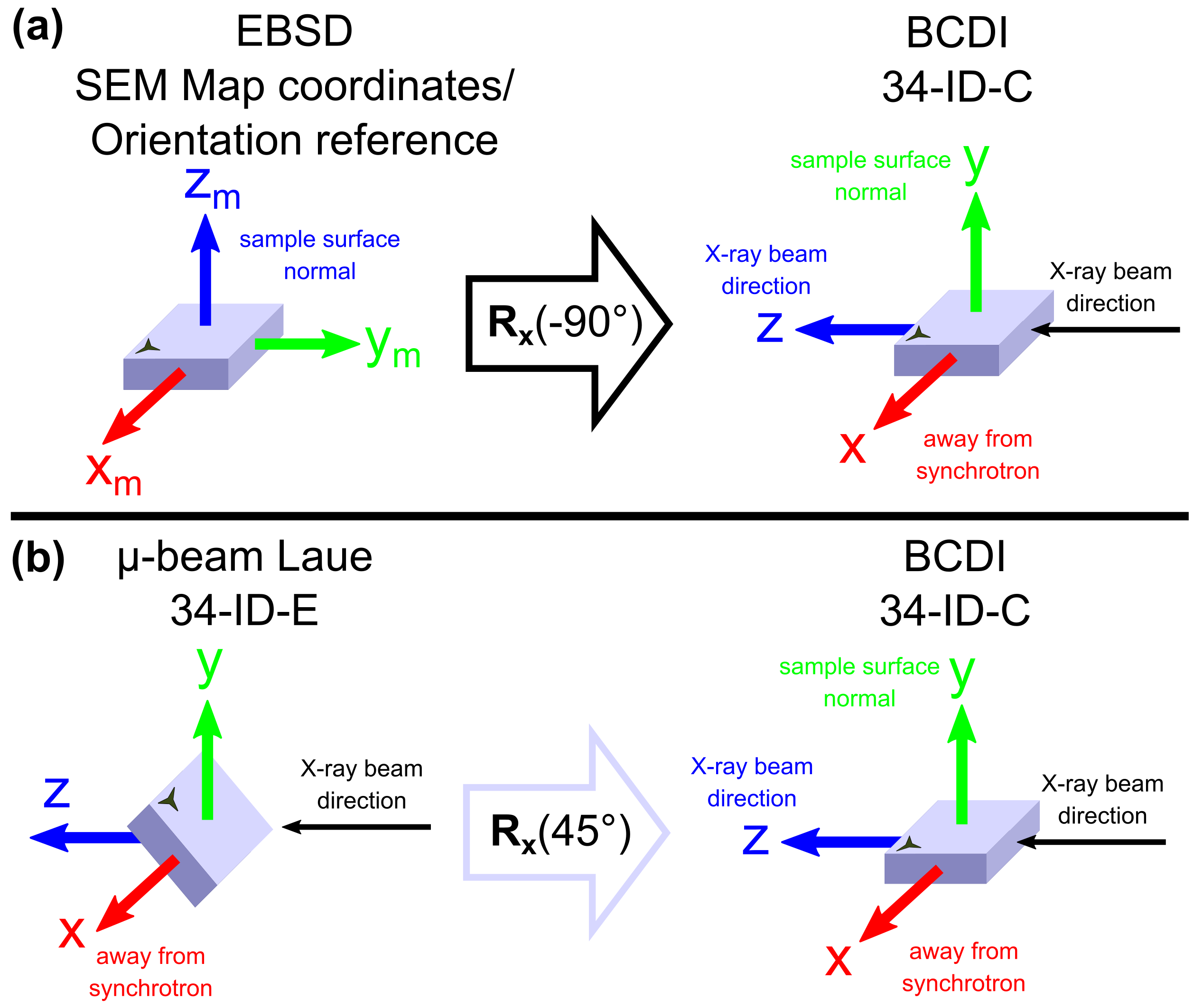}
    \caption{Position of the substrate in the laboratory frame for micro-beam Laue diffraction and EBSD measurements compared to the BCDI laboratory frame ($\mathrm{x}$, $\mathrm{y}$, $\mathrm{z}$). The 3-pointed star represents the orientation of an arbitrary sample feature on the purple substrate. Right-handed rotation matrices are used for the rotations. (a) shows how the EBSD laboratory coordinates, specifically the SEM map coordinates (Fig. \ref{fig:EBSD_orientation}), are related to the BCDI frame. (b) shows the transformation between the laboratory frames for Laue diffraction and BCDI, characterised by a $45\degree$ rotation about the $\mathrm{x}$-axis.}
    \label{fig:Orientation_pic}
\end{figure}

To align the EBSD map frame to the BCDI laboratory frame, a $-90\degree$ rotation about the $\mathrm{x}$-axis is required (Fig. \ref{fig:Orientation_pic}(a)). Combining this with Eq. \ref{eq:U_EBSDm} and Eq. \ref{eq:R_crystal} leads to the formation of the $\mathbf{UB_{34C,\ EBSD}}$ matrix:

\begin{equation} \label{eq:UB_EBSD}
    \mathbf{UB_{34C,\ EBSD}} = \mathbf{R\mathrm{_{x}(-90\degree)}}\mathbf{U_{EBSD,\ m}}\mathbf{R\mathrm{_{crystal}}B}.
\end{equation} 

These matrix-based orientation operators provide a generalised approach to the transformation of orientation, irrespective of the software implementation



\subsection{Micro-beam Laue X-ray diffraction} \label{section:Laue}
Micro-beam Laue diffraction was used to independently verify the lattice orientation of each crystal. This was performed at beamline 34-ID-E at the APS. Further details about the instrument can be found elsewhere \cite{Liu2004,Hofmann2017b}. The sample was positioned with its surface inclined at a $45 \degree$ angle to the incident beam (see Fig. \ref{fig:Orientation_pic}(b)) and diffraction patterns were recorded using a Perkin Elmer flat-panel detector above the sample. 2D fluorescence measurements of the Fe $K\alpha_1$ peak (6.40 keV) were used to identify the spatial position of the crystals, using a monochromatic, 17 keV ($\Delta \lambda/\lambda \approx 10^{-4}$) X-ray beam focused to 0.25 $\mathrm{\mu m}$ x 0.25 $\mathrm{\mu m}$ (h × v) using Kirkpatrick–Baez (KB) mirrors.

\begin{figure}
    \centering
    \includegraphics[height=12cm]{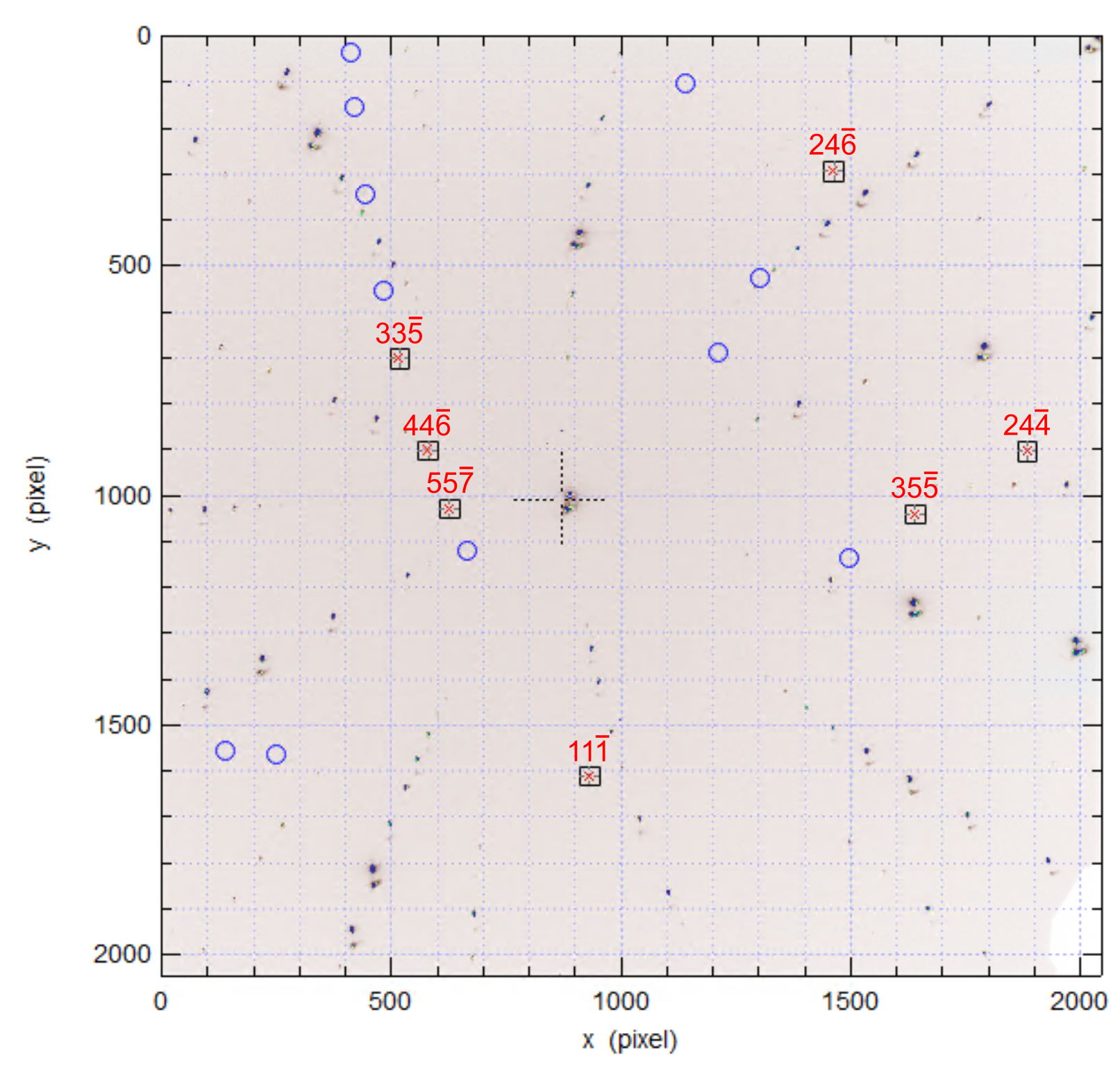}
    \caption{Laue diffraction pattern from microcrystal 1B. The squares show the Bragg peaks that were used for orientation determination, with their corresponding $hkl$ indices shown in red. Weaker reflections from the microcrystal are expected to be inside the blue circles. Other intense peaks in the Laue diffraction pattern belong to the sapphire substrate and are not indexed here for clarity.}
    \label{fig:Laue_pattern}
\end{figure}

Next, a polychromatic X-ray beam was used to collect a Laue diffraction pattern of each crystal (Fig. \ref{fig:Laue_pattern}). The pattern shows weak Bragg reflections from the microcrystals and strong Bragg peaks from the single-crystal sapphire substrate. The two sets of peaks were indexed and fitted using the \textit{LaueGo} software (J. Z. Tischler, tischler@aps.anl.gov). From the indexation, we could determine the $\mathbf{UB}$ matrix (Eq. \ref{eq:UB}). The $\mathbf{UB}$ matrix determined by Laue diffraction at the 34-ID-E instrument is referred to as $\mathbf{UB_{Laue}}$, shown here in Eq. \ref{eq:UB_Laue'},

\begin{equation} \label{eq:UB_Laue'}
    \mathbf{UB_{Laue}} = \begin{bmatrix}
        | & | & | \\
        \mathbf{a^*} & \mathbf{b^*} & \mathbf{c^*} \\
        | & | & | \\
        \end{bmatrix},
\end{equation}

where $\mathbf{a^*}$, $\mathbf{b^*}$ and $\mathbf{c^*}$ are the column (represented by vertical lines) reciprocal space vectors returned by \textit{LaueGo} in units of $\mathrm{nm^{-1}}$ in the 34-ID-E laboratory frame.

To convert Eq. $\mathbf{UB_{Laue}}$ into a $\mathbf{UB}$ matrix for use on the BCDI instrument 34-ID-C, $\mathbf{UB_{34C,\ Laue}}$, the $45 \degree$ rotation of the sample in the Laue laboratory frame must be accounted for (Fig. \ref{fig:Orientation_pic}(b)). To align the micro-beam Laue and BCDI laboratory frames, a rotation of the sample by $45 \degree$ about the $\mathrm{x}$-axis is required \cite{Hofmann2017b}, leading to:

\begin{equation} \label{eq:UB_Laue}
    \mathbf{UB_{34C,\ Laue}} = \mathbf{R\mathrm{_x(45\degree)}}\mathbf{UB_{Laue}}.
\end{equation}


\subsection{Bragg coherent X-ray diffraction imaging} \label{section:BCDI}
BCDI was performed at beamline 34-ID-C at the APS. An \textit{in situ} confocal microscope was used to position the microcrystal within the X-ray beam \cite{Beitra2010}. The sample was illuminated using a 9 keV ($\lambda$ = 0.138 nm) coherent X-ray beam, with a bandwidth of $\Delta\lambda/\lambda \approx 10^{-4}$ from a Si$111$ monochromator. The X-ray beam was focused to a size of 1.1 $\mathrm{\mu m}$ × 1.1 $\mathrm{\mu m}$ (h × v, full width at half-maximum) using KB mirrors. Beam defining slits were used to select the coherent portion of the beam at the entrance to the KB mirrors. For beamline 34-ID-C, the transverse coherence length is $\xi_h > 10\ \mathrm{\mu m}$ and the longitudinal coherence length is $\xi_w\approx0.7\ \mathrm{\mu m}$ at a photon energy of 9 keV \cite{Leake2009}.

The sample needs to be positioned such that a specific $hkl$ Bragg diffraction condition is met to produce a diffraction pattern in the far-field Fraunhofer regime. The orientation matrix determined by EBSD or Laue diffraction is communicated to the \textit{spec} software used on 34-ID-C by defining two reflections that correspond to $hkl$ values associated with laboratory $\mathrm{x}$ (in-plane) and $\mathrm{y}$ (out-of-plane) directions (note the angles referred to here are set elsewhere \cite{Hofmann2017b}):

\begin{enumerate}
    \item The primary reflection (out-of-plane, \textit{y} direction), $\mathbf{\mathbf{H_\perp}}$, where the instrument angles are set to $\delta_{\mathrm{spec}} = 0\degree$, $\gamma_{\mathrm{spec}} = 20\degree$, $\theta_{\mathrm{spec}} = 0\degree$, $\chi_{\mathrm{spec}} = 90\degree$ and $\phi_{\mathrm{spec}} = -10\degree$. The fractional $[h\ k\ l]^\top$ is then:
    
    \begin{equation} \label{eq:primary}
        \begin{bmatrix} h \\ k \\ l \end{bmatrix} = (\mathbf{UB})^{-1}\mathbf{\mathbf{H_\perp}} = (\mathbf{UB})^{-1}\begin{bmatrix} 0 \\ 1 \\ 0 \end{bmatrix}.
    \end{equation}
    
    \item The secondary reflection (in-plane, \textit{x} direction), $\mathbf{\mathbf{H_\parallel}}$, where the instrument angles were set to $\delta_{\mathrm{spec}} = 20\degree$, $\gamma_{\mathrm{spec}} = 0\degree$, $\theta_{\mathrm{spec}} = 10\degree$, $\chi_{\mathrm{spec}} = 90\degree$ and $\phi_{\mathrm{spec}} = 0\degree$. The fractional $[h\ k\ l]^\top$ is then:
    
    \begin{equation} \label{eq:secondary}
        \begin{bmatrix} h \\ k \\ l \end{bmatrix} = (\mathbf{UB})^{-1}\mathbf{\mathbf{H_\parallel}} = (\mathbf{UB})^{-1}\begin{bmatrix} 1 \\ 0 \\ 0 \end{bmatrix}.
    \end{equation}
\end{enumerate}

Here $\mathbf{UB}$ refers to $\mathbf{UB_{34C,\ EBSD}}$ or $\mathbf{UB_{34C,\ Laue}}$. These two fractional $hkl$ vectors are then entered into \textit{spec} as known reflections. Based on this, the expected angular positions of the $\{111\}$ and $\{200\}$ reflections from the sample were calculated. Not all $\{111\}$ and $\{200\}$ reflections could be measured as some may exceed the angular range of the sample and detector motors. Occasionally, manual motor adjustments of a few degrees were required to locate the Bragg peak. Once a Bragg peak was found, the sample tilt and positioning were refined such that the centre of mass was on the centre of the detector positioned $0.5 \mathrm{\ m}$ away from the sample. At this aligned position, the angular and translational positions were saved for each Bragg peak, and were used to determine the true beamline orientation matrix, $\mathbf{UB_{34C}}$, by minimising the least-squares error associated with the measured reflections, 

\begin{equation} \label{eq:UB_beamline}
    \mathrm{min} \sum_{hkl}\Big\|\mathbf{\mathbf{UB_{34C}}}[h\ k\ l]^\top-\mathbf{H_{\mathit{hkl}}}\Big\|^2,
\end{equation}

where $[h\ k\ l]^\top$ are Miller indices in crystal coordinates.

Differences between the true and predicted positions of the reflections using $\mathbf{UB_{34C}}$ arise from a number of different sources. The largest error is the repeatability of sample position in different coordinate systems. The use of Thorlabs 1X1 kinematic mounts, which have angular errors of less than a millidegree, helps with the precise re-mounting of samples. There is also uncertainty in the goniometer precision and alignment with the diffractometer, which can influence the angle readout. Furthermore, the centre of the detector may not be perfectly aligned to the calculated position for a given detector distance or angle. The position of the measured Bragg peak is limited by the energy resolution of the incident X-ray as it affects the Bragg angle. 

Coherent X-ray diffraction patterns (CXDPs) were collected on a 256 × 256 pixel module of a  512 × 512 pixel Timepix area detector (Amsterdam Scientific Instruments) with a GaAs sensor and pixel size of $55 \mathrm{\ \mu m} \times 55 \mathrm{\ \mu m}$ positioned at 1.0 m from the sample to ensure oversampling. CXDPs were recorded by rotating the crystal through an angular range of 0.6° and recording an image every 0.005° with 0.1 s exposure time and 50 accumulations at each angle. 

To optimise the signal to noise ratio and increase the dynamic range of the CXDPs, three repeated scans for each of the $111$, $\bar{1}\bar{1}1$, $200$, $0\bar{2}0$, and $002$ reflections were performed and aligned to maximise their cross-correlation. Once aligned, the minimum acceptable Pearson cross-correlation for summation of CXDPs from a specific Bragg reflection was chosen to be 0.976, similar to previous BCDI studies \cite{Hofmann2018,Hofmann2020}. CXDPs were corrected for dead-time, darkfield, and whitefield prior to cross-correlation alignment. Details regarding the recovery of the real space images using phase retrieval algorithms can be found in Appendix \ref{appendix:phase_retrieval} and the computation of the strain can be found in Appendix \ref{appendix:strain_calculations}.

\subsection{Sample mounting} \label{section:Sample_mounting}
For the SEM, EDX and EBSD analysis, samples were mounted on 12.5 mm diameter SEM specimen pin stubs using silver paint. For micro-beam Laue X-ray diffraction, they were attached to a Thorlabs 1X1 kinematic mount. From here, a Thorlabs kinematic mount adapter between 34-ID-E and 34-ID-C was used to mount the samples for BCDI. This adapter consists of two 1X1 mounts sandwiched together to enable sample orientation to be well-preserved between the beamlines. There is no kinematic mount adapter between the SEM and BCDI instruments. Moreover, the use of magnets in the kinematic mounts inhibits their use for electron microscopy. Across the different instruments, the sample orientation was maintained throughout as shown in Fig. \ref{fig:Orientation_pic}, which has an arbitrary sample feature to illustrate the respective orientations.

\section{Results and Discussion} \label{section:results_and_discussion}
\subsection{Orientation matrix comparison}
The angular mismatch between two $\mathbf{UB}$ matrices,$\mathbf{UB_1}$ and $\mathbf{UB_2}$, can be determined by converting $\mathbf{UB_1(UB_2)^{-1}}$ into a rotation vector. A rotation matrix, $\mathbf{R}$, can be converted using Rodrigues' rotation formula in matrix exponential form,

\begin{equation} \label{eq:Rodrigues}
    \mathbf{R} = e^{\mathbf{w_m}},
\end{equation}

where $\mathbf{w_m}$ is an antisymmetric matrix,

\begin{equation} \label{eq:Antisymmetric}
    \mathbf{w_m} = \begin{bmatrix}
        0 & -w_z & w_y \\
        w_z & 0 & -w_x \\
        -w_y & w_x & 0 \\
        \end{bmatrix},
\end{equation}

which contains the elements of the rotation vector $\mathbf{w}=[w_x, w_y, w_z]^\top$. The rotation vector is defined by a rotation axis $\hat{\mathbf{w}}$ multiplied by a rotation, $\theta$. If the two orientation matrices are different, $\mathbf{UB_1(UB_2)}^{-1}$ can be converted to a rotation vector where the angular mismatch is $\theta$. If $\mathbf{UB_1} = \mathbf{UB_2}$, then $\mathbf{UB_1(UB_2)^{-1}} = I_3$ and therefore $\theta=0$.

Here we set $\mathbf{UB_1}$ and $\mathbf{UB_2}$ as $\mathbf{UB_{34C,\ EBSD}}$, $\mathbf{UB_{34C,\ Laue}}$ or $\mathbf{UB_{34C}}$. First we rearrange Eq. \ref{eq:Rodrigues} to calculate $\mathbf{w_m}$:

\begin{equation} \label{eq:log}
    \mathbf{w_m} = \mathrm{log}[\mathbf{UB_1(UB_2)}^{-1}],
\end{equation}

where $\mathrm{log}$ here refers to the matrix natural logarithm. Next we reconstruct the rotation vector using Eq. \ref{eq:Antisymmetric} and calculate its magnitude to obtain the mismatch (Eq. \ref{eq:mismatch}),

\begin{equation} \label{eq:mismatch}
    \theta = \|\mathbf{w}\|= \|[\mathbf{w_m}(3,2),\mathbf{w_m}(1,3),\mathbf{w_m}(2,1)]^\top\|.
\end{equation}

To calculate the angular mismatch between $\mathbf{UB_{34C,\ EBSD}}$ and other orientation matrices, the permutation of $n_{x,y,z}$ that produced the smallest error was chosen. Tables \ref{table:Angle_mismatch1} - \ref{table:Angle_mismatch3} show the angular mismatch of $\mathbf{UB_{34C,\ Laue}}$ and $\mathbf{UB_{34C,\ EBSD}}$ compared to $\mathbf{UB_{34C}}$ for substrates 1 - 3. 

\begin{table}
    \begin{center}
        \caption{$\mathrm{Angular\ differences\ (\degree)\ between\ \mathbf{UB_{34C}},\ \mathbf{UB_{34C,\ Laue}},\ and\ \mathbf{UB_{34C,\ EBSD}}\ for\ substrate\ 1.}$}
        \begin{tabular}{l|ccc}      
            \hline\multicolumn{1}{c|}{1B} & $\mathbf{UB_{34C}}$ & $\mathbf{UB_{34C,\ Laue}}$ & $\mathbf{UB_{34C,\ EBSD}}$ \\ \hline
            $\mathbf{UB_{34C}}$ & - & - & - \\
            $\mathbf{UB_{34C,\ Laue}}$ & 9.72 & - & - \\
            $\mathbf{UB_{34C,\ EBSD}}$ & 11.0 & 2.37 & - \\ \\ \hline
            \multicolumn{1}{c|}{1C} & $\mathbf{UB_{34C}}$ & $\mathbf{UB_{34C,\ Laue}}$ & $\mathbf{UB_{34C,\ EBSD}}$ \\ \hline
            $\mathbf{UB_{34C}}$ & - & - & - \\
            $\mathbf{UB_{34C,\ Laue}}$ & 2.22 & - & - \\
            $\mathbf{UB_{34C,\ EBSD}}$ & 5.35 & 3.28 & - \\ \\ \hline
            \multicolumn{1}{c|}{1E} & $\mathbf{UB_{34C}}$ & $\mathbf{UB_{34C,\ Laue}}$ & $\mathbf{UB_{34C,\ EBSD}}$ \\ \hline
            $\mathbf{UB_{34C}}$ & - & - & - \\
            $\mathbf{UB_{34C,\ Laue}}$ & 9.70 & - & - \\
            $\mathbf{UB_{34C,\ EBSD}}$ & 11.0 & 1.97 & - \\ \\ \hline
            \multicolumn{1}{c|}{1F} & $\mathbf{UB_{34C}}$ & $\mathbf{UB_{34C,\ Laue}}$ & $\mathbf{UB_{34C,\ EBSD}}$ \\ \hline
            $\mathbf{UB_{34C}}$ & - & - & - \\
            $\mathbf{UB_{34C,\ Laue}}$ & 7.14 & - & - \\
            $\mathbf{UB_{34C,\ EBSD}}$ & 9.56 & 3.05 & - \\ \\ \hline
            \multicolumn{1}{c|}{Average} & $\mathbf{UB_{34C}}$ & $\mathbf{UB_{34C,\ Laue}}$ & $\mathbf{UB_{34C,\ EBSD}}$ \\ \hline
            $\mathbf{UB_{34C}}$ & - & - & - \\
            $\mathbf{UB_{34C,\ Laue}}$ & 7.19 & - & - \\
            $\mathbf{UB_{34C,\ EBSD}}$ & 9.23 & 2.67 & - \\
        \end{tabular}
        \label{table:Angle_mismatch1}
    \end{center}
\end{table}

\begin{table}
    \begin{center}
    \caption{$\mathrm{Angular\ differences\ (\degree)\ between\ \mathbf{UB_{34C}},\ \mathbf{UB_{34C,\ Laue}},\ and\ \mathbf{UB_{34C,\ EBSD}}\ for\ substrate\ 2.}$}
        \begin{tabular}{l|ccc}      
            \hline\multicolumn{1}{c|}{2A} & $\mathbf{UB_{34C}}$ & $\mathbf{UB_{34C,\ Laue}}$ & $\mathbf{UB_{34C,\ EBSD}}$ \\ \hline
            $\mathbf{UB_{34C}}$ & - & - & - \\
            $\mathbf{UB_{34C,\ Laue}}$ & 1.81 & - & - \\
            $\mathbf{UB_{34C,\ EBSD}}$ & 2.00 & 0.845 & - \\ \\ \hline
            \multicolumn{1}{c|}{2B} & $\mathbf{UB_{34C}}$ & $\mathbf{UB_{34C,\ Laue}}$ & $\mathbf{UB_{34C,\ EBSD}}$ \\ \hline
            $\mathbf{UB_{34C}}$ & - & - & - \\
            $\mathbf{UB_{34C,\ Laue}}$ & 0.366 & - & - \\
            $\mathbf{UB_{34C,\ EBSD}}$ & 3.33 & 3.56 & - \\ \\ \hline
            \multicolumn{1}{c|}{2C} & $\mathbf{UB_{34C}}$ & $\mathbf{UB_{34C,\ Laue}}$ & $\mathbf{UB_{34C,\ EBSD}}$ \\ \hline
            $\mathbf{UB_{34C}}$ & - & - & - \\
            $\mathbf{UB_{34C,\ Laue}}$ & 11.0 & - & - \\
            $\mathbf{UB_{34C,\ EBSD}}$ & 11.2 & 0.306 & - \\ \\ \hline
            \multicolumn{1}{c|}{2D} & $\mathbf{UB_{34C}}$ & $\mathbf{UB_{34C,\ Laue}}$ & $\mathbf{UB_{34C,\ EBSD}}$ \\ \hline
            $\mathbf{UB_{34C}}$ & - & - & - \\
            $\mathbf{UB_{34C,\ Laue}}$ & 0.392 & - & - \\
            $\mathbf{UB_{34C,\ EBSD}}$ & 0.659 & 0.269 & - \\ \\ \hline
            \multicolumn{1}{c|}{2E} & $\mathbf{UB_{34C}}$ & $\mathbf{UB_{34C,\ Laue}}$ & $\mathbf{UB_{34C,\ EBSD}}$ \\ \hline
            $\mathbf{UB_{34C}}$ & - & - & - \\
            $\mathbf{UB_{34C,\ Laue}}$ & 0.977 & - & - \\
            $\mathbf{UB_{34C,\ EBSD}}$ & 4.34 & 4.11 & - \\ \\ \hline
            \multicolumn{1}{c|}{2F} & $\mathbf{UB_{34C}}$ & $\mathbf{UB_{34C,\ Laue}}$ & $\mathbf{UB_{34C,\ EBSD}}$ \\ \hline
            $\mathbf{UB_{34C}}$ & - & - & - \\
            $\mathbf{UB_{34C,\ Laue}}$ & 0.425 & - & - \\
            $\mathbf{UB_{34C,\ EBSD}}$ & 1.58 & 1.23 & - \\ \\ \hline
            \multicolumn{1}{c|}{Average} & $\mathbf{UB_{34C}}$ & $\mathbf{UB_{34C,\ Laue}}$ & $\mathbf{UB_{34C,\ EBSD}}$ \\ \hline
            $\mathbf{UB_{34C}}$ & - & - & - \\
            $\mathbf{UB_{34C,\ Laue}}$ & 2.49 & - & - \\
            $\mathbf{UB_{34C,\ EBSD}}$ & 3.86 & 1.52 & - \\
        \end{tabular}
        \label{table:Angle_mismatch2}
    \end{center}
\end{table}

\begin{table}
    \caption{$\mathrm{Angular\ differences\ (\degree)\ between\ \mathbf{UB_{34C}},\ \mathbf{UB_{34C,\ Laue}},\ and\ \mathbf{UB_{34C,\ EBSD}}\ for\ substrate\ 3.}$}
    \begin{center}
        \begin{tabular}{l|ccc}      
            \hline\multicolumn{1}{c|}{3A} & $\mathbf{UB_{34C}}$ & $\mathbf{UB_{34C,\ Laue}}$ & $\mathbf{UB_{34C,\ EBSD}}$ \\ \hline
            $\mathbf{UB_{34C}}$ & - & - & - \\
            $\mathbf{UB_{34C,\ Laue}}$ & 0.369 & - & - \\
            $\mathbf{UB_{34C,\ EBSD}}$ & 2.28 & 2.25 & - \\ \\ \hline
            \multicolumn{1}{c|}{3B} & $\mathbf{UB_{34C}}$ & $\mathbf{UB_{34C,\ Laue}}$ & $\mathbf{UB_{34C,\ EBSD}}$ \\ \hline
            $\mathbf{UB_{34C}}$ & - & - & - \\
            $\mathbf{UB_{34C,\ Laue}}$ & 9.69 & - & - \\
            $\mathbf{UB_{34C,\ EBSD}}$ & 10.7 & 2.09 & - \\ \\ \hline
            \multicolumn{1}{c|}{Average} & $\mathbf{UB_{34C}}$ & $\mathbf{UB_{34C,\ Laue}}$ & $\mathbf{UB_{34C,\ EBSD}}$ \\ \hline
            $\mathbf{UB_{34C}}$ & - & - & - \\
            $\mathbf{UB_{34C,\ Laue}}$ & 5.03 & - & - \\
            $\mathbf{UB_{34C,\ EBSD}}$ & 6.49 & 1.74 & - \\
        \end{tabular}
        \label{table:Angle_mismatch3}
    \end{center}
\end{table}

The average angular mismatches for all crystals between orientation matrices: $\mathbf{UB_{34C,\ Laue}}$ and $\mathbf{UB_{34C}}$ is 4.48\degree, $\mathbf{UB_{34C,\ EBSD}}$ and $\mathbf{UB_{34C}}$ is 6.09\degree, and $\mathbf{UB_{34C,\ Laue}}$ and $\mathbf{UB_{34C,\ EBSD}}$ is 1.95\degree. Generally, $\mathbf{UB_{34C,\ Laue}}$ is most similar to $\mathbf{UB_{34C}}$. This is expected as there is a Thorlabs kinematic mount adapter between 34-ID-E and 34-ID-C for the precise angular alignment samples. A larger difference is observed when comparing $\mathbf{UB_{34C,\ EBSD}}$ to $\mathbf{UB_{34C}}$. This is due to the manual removal of the SEM pin stub from the electron microscope, which then needs to be re-secured to the kinematic mount. The cylindrical pin permits a greater degree of rotational freedom, thereby increasing the angular mismatch when $\mathbf{UB_{34C,\ EBSD}}$ is considered. Despite this increased angular freedom in the pin, a $2\degree$ increase in the angular uncertainty when using $\mathbf{UB_{34C,\ EBSD}}$ instead of $\mathbf{UB_{34C,\ Laue}}$ is still a very accurate result. This means only a slightly larger angular range needs to be explored in alignment. 

The alignment of crystals for BCDI using EBSD remains much more time efficient because micro-beam Laue diffraction pre-alignment at 34-ID-E is no longer required. The use of EBSD for pre-alignment of BCDI samples also affords greater flexibility in experiment type. The ability to pre-characterise samples off-site should substantially increase throughput and make MBCDI a much more widely accessible technique, especially at beamlines without pink beam capability or access to a nearby Laue instrument.

\subsection{Determination of strain}
\begin{figure}
    \centering
    \includegraphics[height=14cm]{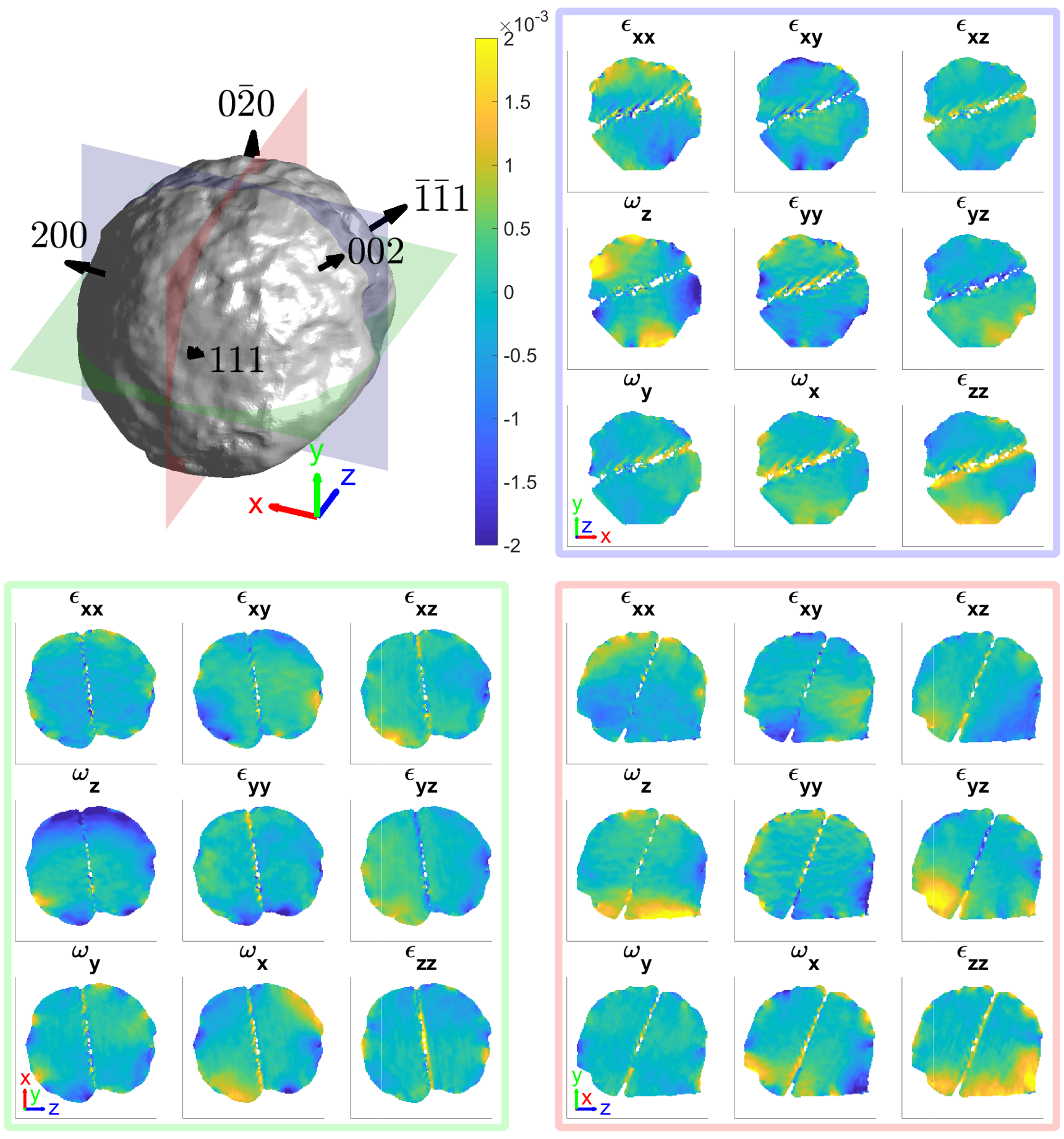}
    \caption{Average morphology of the $111$, $\bar{1}\bar{1}1$, $200$, $0\bar{2}0$, and $002$ reflections for Fe-Ni crystal 1B. The region of missing intensity in the middle of the crystal corresponds to a twinned region of the crystal, which is only visible in the $111$ reconstruction (Fig. \ref{fig:Morphology_refl} in Appendix \ref{appendix:phase_retrieval}). The average morphology taken over all five reflections is based on an amplitude threshold of 0.30. The slices through the strain and rotation tensor components using the average morphology are shown for the planes indicated at $x = 2.5\ \mathrm{nm}$ (red), $y = 2.5\ \mathrm{nm}$ (green), and $z = 2.5\ \mathrm{nm}$ (blue) from the centre of mass of the microcrystal. The amplitude threshold is 0.30 and the coordinate axes are 100 nm long. Supplementary video (SV) 1-3 show the strain and rotation tensor components throughout the volume along the $\mathrm{x}$, $\mathrm{y}$, and $\mathrm{z}$ axis respectively.}
    \label{fig:Tensor}
\end{figure}

MBCDI allows the strain and rotation tensors to be computed if at least three reflections are measured, providing more information about the crystal defects present \cite{Hofmann2020,Hofmann2017a,Hofmann2018,Phillips2020}. Fig. \ref{fig:Tensor} shows the strain and rotation tensors reconstructed from five measured Bragg reflections of crystal 1B. The $\mathbf{\epsilon}_{xx}$, $\mathbf{\epsilon}_{yy}$, and $\mathbf{\epsilon}_{zz}$ slices show defects close to the edge that may not be resolved if analysing a single reflection alone \cite{Yang2021} as some crystal defects, such as dislocations, are visible only when $\mathbf{Q_\mathit{hkl}}\mathbf{\cdot b}\neq0$, where $\mathbf{b}$ is the Burgers vector \cite{Williams2009}.

These results were produced using a refined method for the computation of the strain and rotation tensor. A general approach for the computation of both these tensors is described in Appendix \ref{appendix:strain_calculations}. This relies on the recovery of the phase of the CXDP. The intensity of the CXDP, which is the squared magnitude of the Fourier transform, $\mathcal{F}$, of the complex crystal electron density, $f$. The solution for the recovered phase is non-unique, as global phase offsets, $C$, can produce the same CXDP, i.e. $|\mathcal{F}[f(\psi(\mathbf{r}))]|^2= |\mathcal{F}[f(\psi(\mathbf{r})+C)]|^2$. After phase retrieval, the phase values are bound between $[-\pi, \pi]$, which describes the periodic nature of the lattice structure, but not necessarily the true complex crystal electron density. For instance, if the projected displacement in the direction of $\mathbf{Q_\mathit{hkl}}$ is greater than $\pi/|\mathbf{Q_\mathit{hkl}}|$, then a phase jump, where the phase difference is $2\pi$ between two pixels, will occur. These phase jumps cause discontinuities in the derivatives of the phase, $\partial \psi_{hkl}(\mathbf{r})/\partial j$, where $j$ corresponds to the spatial $\mathrm{x}$, $\mathrm{y}$, or $\mathrm{z}$ coordinate, leading to spurious, large strains. Typically, phase unwrapping algorithms can be used to remove phase jumps, but dislocations have characteristic phase vortices \cite{Clark2015} that end at dislocation lines, meaning that phase jumps associated with dislocations cannot be unwrapped.

To account for this, Hofmann \textit{et al.} \cite{Hofmann2020} demonstrated an approach by producing two additional copies of the phase with phase offsets of $-\frac{\pi}{2}$ and $\frac{\pi}{2}$ respectively. This shifts phase jumps to different locations, and by choosing the phase gradient with the smallest magnitude for each voxel, the correct phase derivatives can be found. Here, we employ a more efficient method used in coherent X-ray diffraction tomography from Guizar-Sicairos \textit{et al.} \cite{Guizar-Sicairos2011}, which has also been applied in the Bragg geometry using ptychography \cite{Li2021b}. Rather than making multiple copies of the phase, we take the derivative of the complex exponential of the phase and determine the phase gradient using the chain rule:

\begin{equation} \label{eq:chain_rule}
    \frac{\partial\psi_{hkl}(\mathbf{r})}{\partial j} = \mathrm{Re}\left(\frac{\partial e^{i\psi_{hkl}(\mathbf{r})}}{\partial j}\bigg/ie^{i\psi_{hkl}}\right).
\end{equation}

\begin{figure}
    \centering
    \includegraphics[height=5cm]{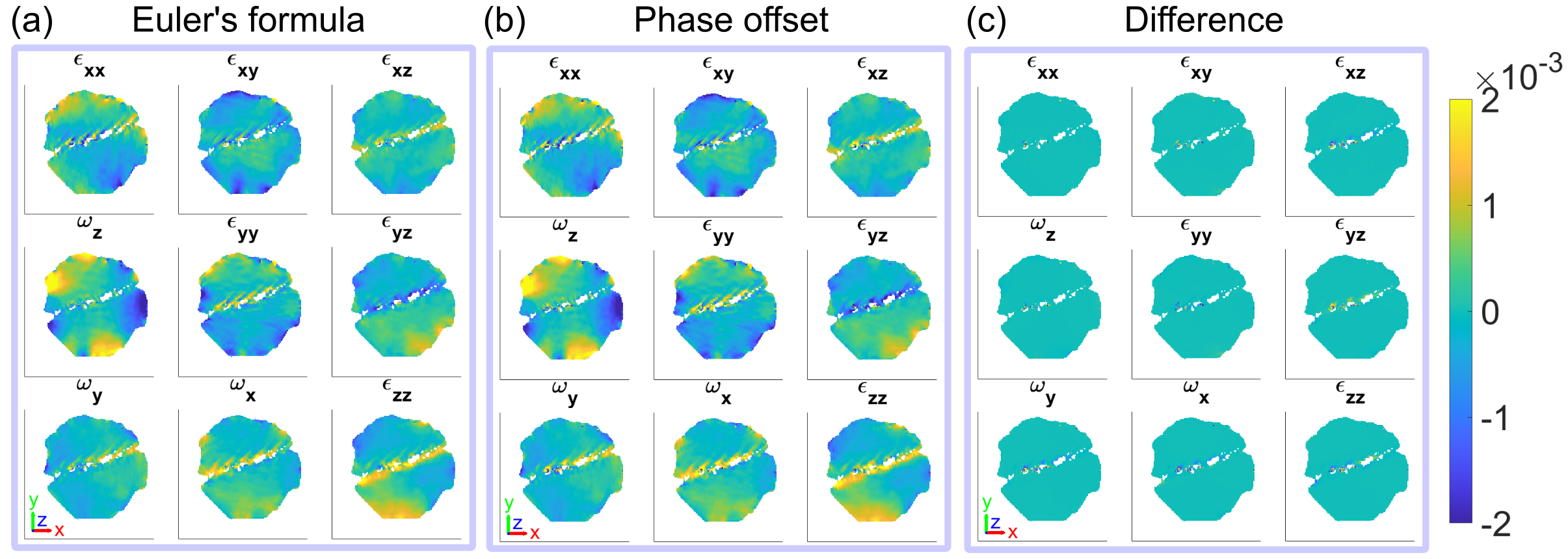}
    \caption{A comparison of the strain and rotation tensors at $z = 2.5\ \mathrm{nm}$, as shown in Fig. \ref{fig:Tensor}, computed using two different methods. (a) shows the strain and rotation tensors as computed using Eq. \ref{eq:chain_rule} \cite{Guizar-Sicairos2011}. (b) shows the strain and rotation tensors as computed by introducing phase offsets and choosing the phase gradient with the minimum value \cite{Hofmann2020}. (c) shows the difference between the results. The amplitude threshold for the reconstructions is 0.30 and the magnitude of the coordinate axes is 100 nm.}
    \label{fig:Tensor_difference}
\end{figure}

Here $e^{i\psi_{hkl}(\mathbf{r})}$ is a circle expressed using Euler's formula, where the phase jumps disappear. Fig. \ref{fig:Tensor_difference} shows the difference in the two methods for computing the lattice rotation and strain tensors. We can see that the procedure based on phase offsets \cite{Hofmann2020} fails to fully resolve the details in the regions with high strain, i.e. around the edges and central region of missing intensity. The new approach of computing phase gradients successfully deals with these complex regions.

\begin{figure}
    \centering
    \includegraphics[height=7cm]{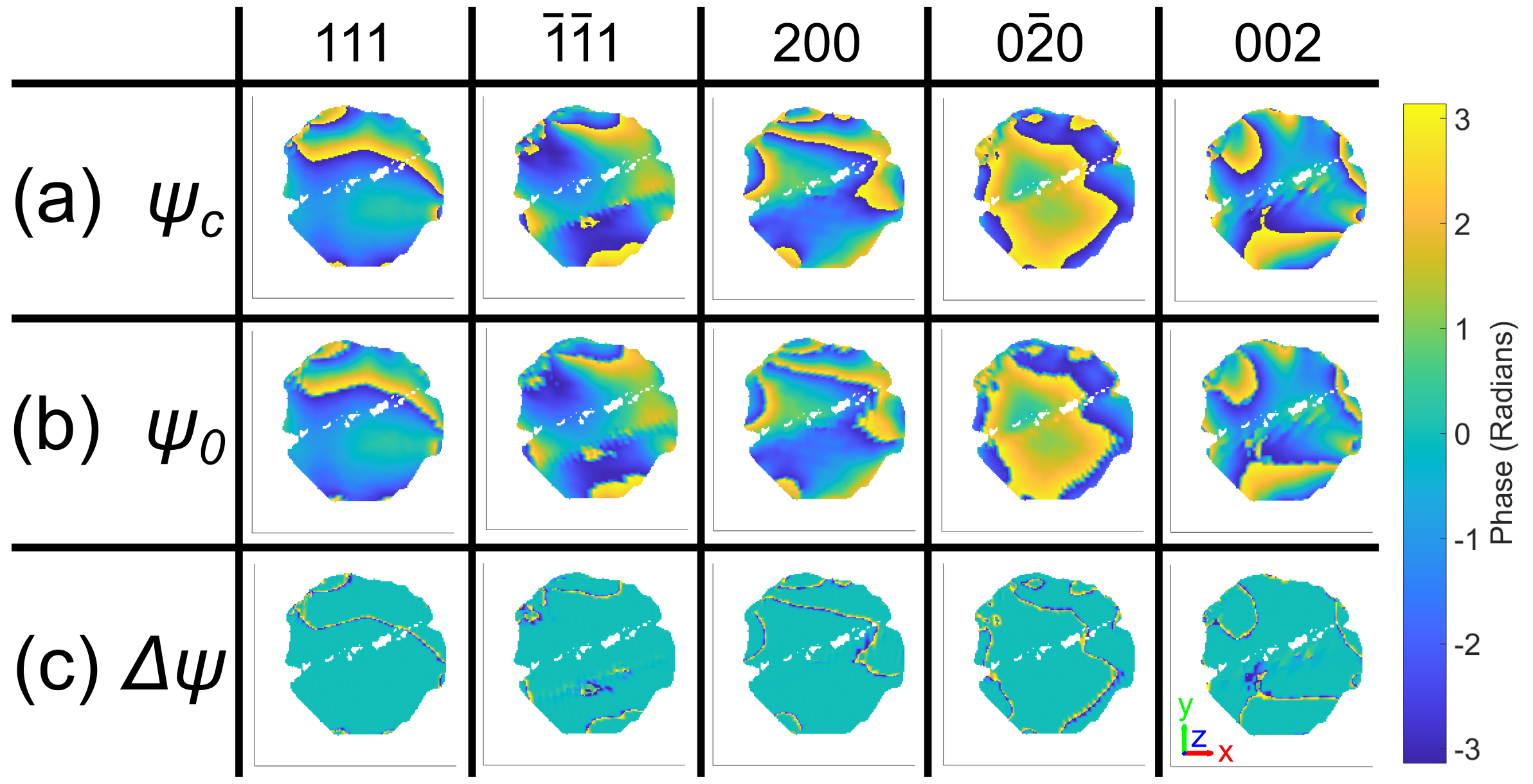}
    \caption{Phases for the $111$, $\bar{1}\bar{1}1$, $200$, $0\bar{2}0$, and $002$ reflections at $z = 2.5\ \mathrm{nm}$. We compare the new method of interpolating the complex quantity of the phase, $\psi_c$, (a), to interpolating the phase alone, $\psi_0$, (b). The difference between $\psi_c$ and $\psi_0$, $\Delta\psi$, highlights the information that can only be resolved around phase jumps in $\psi_c$ (c). The amplitude threshold for the reconstructions is 0.30 and the size of the coordinate axes is 100 nm. Here the average morphology was used, hence the $111$ reconstruction shows a region of missing intensity corresponding to a twin, which is not present in the original $111$ reconstructed morphology (Fig. \ref{fig:Morphology_refl} in Appendix \ref{appendix:phase_retrieval}).}
    \label{fig:Phase_difference}
\end{figure}

Furthermore, we apply this to the interpolation of the phase from detector conjugated space to sample space, by interpolating the complex quantity, $e^{i\psi_{hkl}}$, instead of $\psi_{hkl}$. This avoids the blurring of phase jumps that occurs during direct interpolation of $\psi_{hkl}$, shown in Fig. \ref{fig:Phase_difference}.

This refinement of strain and rotation tensor computation allows for a more accurate reconstruction of crystal defects and their associated nanoscale lattice strains. It also decreases the time required to compute the tensors since phase offsets need to be applied before and after mapping the crystal from detector conjugated space to orthogonal sample space. This will play an important role in the analysis of large MBCDI datasets, such as those obtained from \textit{in situ} or \textit{operando} experiments that reveal crystal defects interacting with their environment. Reconstruction accuracy in multi-Bragg peak phase retrieval algorithms that involve a shared displacement field constraint applied to all reflections would also be improved by implementing this more accurate interpolation of phase values \cite{Newton2020,Gao2021,Wilkin2021}. 


\section{Conclusion} \label{section:conclusion}
We demonstrate that the orientation of various microcrystals on different substrates can be found using EBSD and used to align BCDI experiments. Results indicate a $\sim2\degree$ increase in angular error when using EBSD alignment compared to Laue diffraction alignment, which is still within reasonable tolerance for the search for Bragg peaks. Importantly, using EBSD to pre-align crystals allows beamtime to be more effectively utilised for BCDI dataset collection. It also removes the need for BCDI and Laue instrument coordination and enables MBCDI on BCDI instruments that do not have pink beam capability or a Laue instrument nearby. Using the orientation matrix obtained from EBSD, five reflections are located on a Fe-Ni microcrystal and full 3D strain and rotation tensors are recovered. When computing the tensors, we demonstrate a more efficient approach to resolving phase jumps, by implementing a complex phase quantity to calculate and interpolate the phase. This allows for the phase to be unwrapped and the correct strain to be resolved in the vicinity of dislocations. These refinements make BCDI a more accessible microscopy tool.
     

\section{Data availability}
The processed diffraction patterns, final reconstructions, and data analysis scripts, including a script to compute the orientation matrix using EBSD, are publicly available at 10.5281/zenodo.6383408. 

\pagebreak

\appendix

\section{Phase retrieval} \label{appendix:phase_retrieval}
The reconstruction process of CXDPs was done independently for each reflection, in two stages, using the output from the previous stage to seed the next phasing stage (listed below). The CXDPs have a size of $256 \times 256 \times 128$ voxels.
\begin{enumerate}
    \item Each reconstruction was seeded with a random guess. A guided phasing approach \cite{Chen2007} with 40 individuals and four generations was used with a geometric average breeding mode. For each generation and population, a block of 20 error reduction (ER) and 180 hybrid input-output (HIO) iterations, with $\beta = 0.9$, was repeated three times. This was followed by 20 ER iterations to return the final object. The shrinkwrap algorithm \cite{Marchesini2003} with a threshold of 0.1 was used to update the real-space support every iteration. The $\sigma$ for the Gaussian kernel shrinkwrap smoothing for each generation was $\sigma = 2.0, 1.5, 1.0, 1.0$ respectively. The best reconstruction was determined using a sharpness criterion, as it is the most appropriate metric for crystals containing defects \cite{Ulvestad2017b}. The two worst reconstructions were removed after each generation.
    
    \item The reconstruction was seeded with the output from stage 1. This stage was identical to stage 1, except now for each generation and population, a block of 20 error reduction (ER) and 180 hybrid input-output (HIO) iterations, with $\beta = 0.9$, was repeated 15 times, followed by 1000 ER iterations. Here the final 50 iterates were averaged to produce the final image.
\end{enumerate}

\begin{figure} 
    \centering
    \includegraphics[height=9cm]{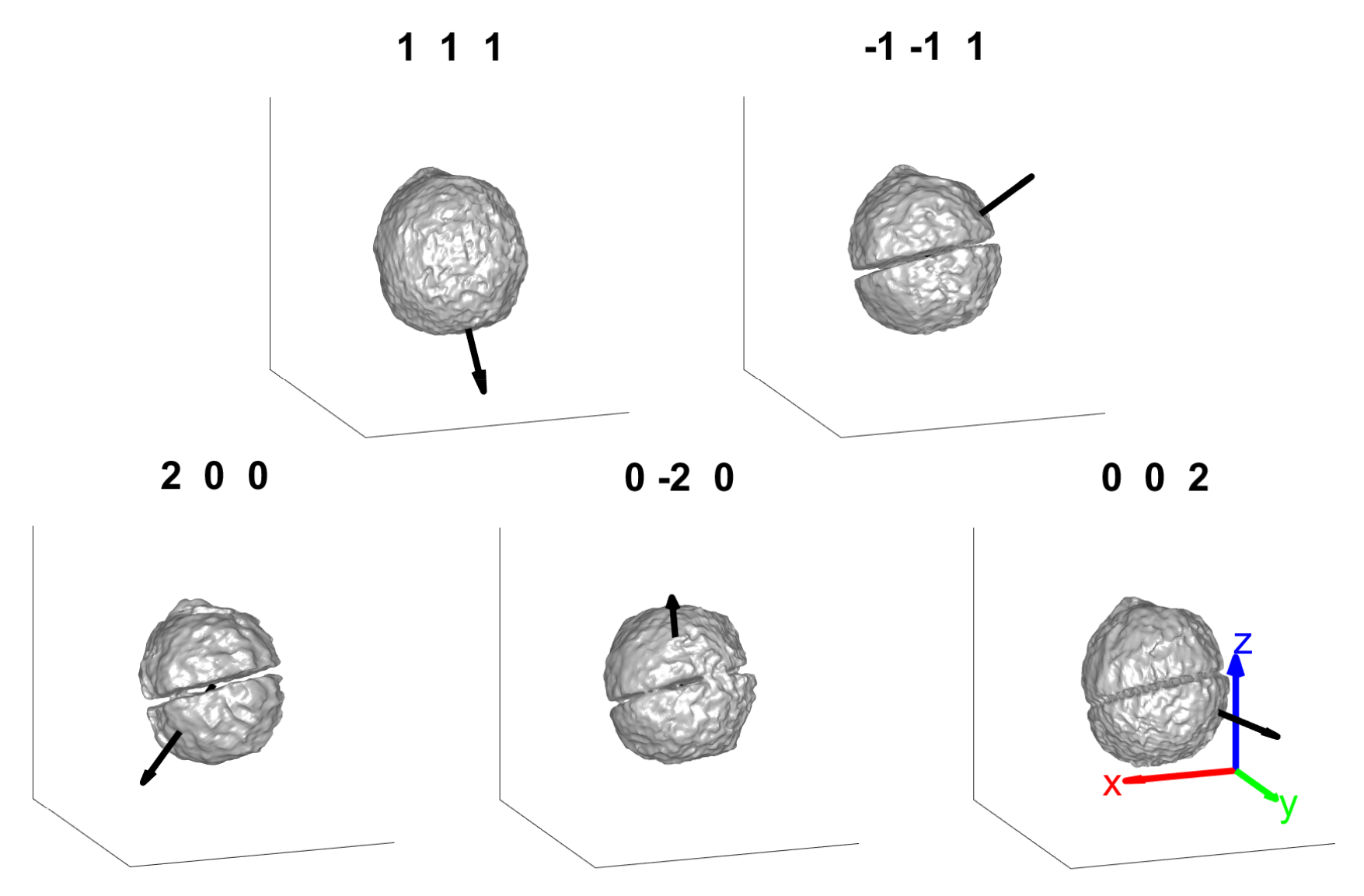}
    \caption{The sample morphology for each measured reflection labelled with the scattering vector. The average morphology for the five reflections is shown in Fig. \ref{fig:Tensor}. The amplitude threshold for the reconstructions is 0.30 and the size of the coordinate axes is 500 nm.}
    \label{fig:Morphology_refl}
\end{figure}

The overall average 3D spatial resolution was 35 nm. This was determined by differentiating the electron density amplitude across the crystal/air interface for the five reflection directions and fitting a Gaussian to each of the profiles. The reported spatial resolution is the averaged full width at half maximum of the Gaussian profiles. The average 3D spatial resolution was 30, 38, 39, 31 and 39 nm at for the $111$, $\bar{1}\bar{1}1$, $200$, $0\bar{2}0$, and $002$ reconstructions (Fig. \ref{fig:Morphology_refl}) respectively.

\section{Strain and rotation tensor calculations}
\label{appendix:strain_calculations}
The full 3D lattice strain tensor, $\mathbf{\varepsilon}(\mathbf{r})$, and rotation tensor, $\mathbf{\omega}(\mathbf{r})$, are given by \cite{Constantinescu2008}:

\begin{align}
    \begin{split} \label{eq:lattice_strain_tensor}
        \mathbf{\varepsilon}(\mathbf{r}) &= \frac{1}{2}\left\{\nabla \mathbf{u}(\mathbf{r})+[\nabla\mathbf{u}(\mathbf{r})]^\top\right\}\mathrm{,\ and}\\
    \end{split}\\\nonumber\\
    \begin{split} \label{eq:rotation_tensor}
        \mathbf{\omega}(\mathbf{r}) &= \frac{1}{2}\left\{\nabla \mathbf{u}(\mathbf{r})-[\nabla\mathbf{u}(\mathbf{r})]^\top\right\}\\
    \end{split}
\end{align}

which rely on the reconstruction of $\mathbf{u}(\mathbf{r})$. With a single BCDI measurement, we can determine one component of the displacement field. For multi-reflection BDCI (MBCDI), if at least three linearly independent reflections are measured, $\mathbf{u}(\mathbf{r})$ can be determined by minimising the least-squares error \cite{Hofmann2017b,Newton2010},

\begin{equation} \label{eq:least-squares_displacement}
    E(\mathbf{r}) = \sum_{hkl}[\mathbf{Q}_{hkl}\cdot\mathbf{u}(\mathbf{r})-\psi_{hkl}(\mathbf{r})]^2,
\end{equation}

for every voxel in the sample. Here, we use the modified approach by Hofmann \textit{et al.} \cite{Hofmann2020}, in which case, the squared error between phase gradients is minimised:

\begin{equation} \label{eq:least-squares_gradient}
    E(\mathbf{r})_j = \sum_{hkl,j} \left[\mathbf{Q}_{hkl}\cdot\frac{\partial \mathbf{u}(\mathbf{r})}{\partial j}-\frac{\partial \psi_{hkl}(\mathbf{r})}{\partial j}\right]^2,
\end{equation}

where $j$ corresponds to the spatial $\mathrm{x}$, $\mathrm{y}$, or $\mathrm{z}$ coordinate, to find $\nabla \mathbf{u}(\mathbf{r})$ directly for the computation of $\mathbf{\varepsilon}(\mathbf{r})$, and $\mathbf{\omega}(\mathbf{r})$ in Eqs. \ref{eq:lattice_strain_tensor} and \ref{eq:rotation_tensor} respectively.

In Fig. \ref{fig:Tensor}, five linearly independent reflections were measured and reconstructed to assemble the strain and rotation tensor. To assess the reliability of the MBCDI measurements and phase retrieval procedure, we calculate the strain tensor using four out of the five reflections to predict the strain projected along the scattering vector of the fifth reflection. First, the phase gradients for each reflection are calculated from the strain tensor components,

\begin{equation} \label{ch5:eq:phase_grad}
    \frac{\partial\psi_{hkl}(\mathbf{r})}{\partial j} = \mathbf{Q}_{hkl}\cdot\frac{\partial \mathbf{u}(\mathbf{r})}{\partial j},
\end{equation}

and then used to calculate the strain fields projected along the scattering vector, $\epsilon_{hkl}\mathbf{(r)}$:

\begin{equation} \label{ch5:eq:strain}
    \epsilon_{hkl}\mathbf{(r)} = \nabla\mathbf{\psi_\mathit{hkl}(r)}\cdot\frac{\mathbf{Q_\mathit{hkl}}}{|\mathbf{Q_\mathit{hkl}}|^2}.
\end{equation}

This is compared to the strain field projected along the scattering vector computed from the measured phase gradients using the approach presented in the main text (Eq. \ref{eq:chain_rule}). 

\begin{figure} 
    \centering
    \includegraphics[height=7cm]{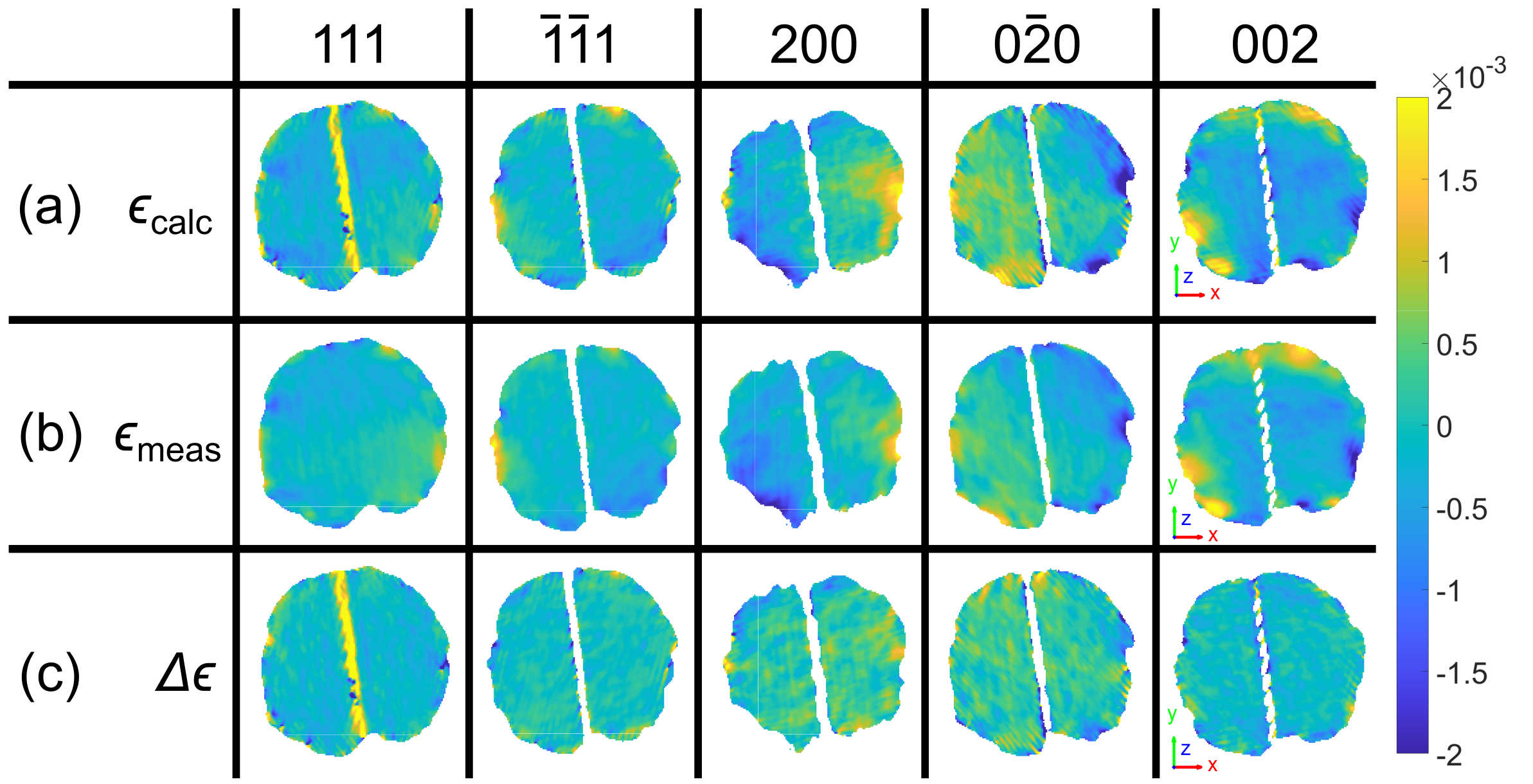}
    \caption{A comparison between the (a) calculated and (b) measured strain for the $111$, $\bar{1}\bar{1}1$, $200$, $0\bar{2}0$, and $002$ reflections at $y = 2.5\ \mathrm{nm}$. The calculated strain is computed using the strain tensor determined using the other four reflections following the methodology presented in article. (c) shows the measured strain subtracted from the calculated strain. The amplitude threshold for the reconstructions is 0.30 and the size of the coordinate axes is 100 nm.}
    \label{fig:Strain_error}
\end{figure}

The average strain error for each reflection is computed by summing the magnitude of the difference between the calculated and measured strain and dividing the sum by the number of voxels in the morphology of each reconstruction (Fig. \ref{fig:Strain_error}). The average strain error for each reflection, listed in parentheses, is $3.5\times10^{-4}$ ($111$), $2.1\times10^{-4}$ ($\bar{1}\bar{1}1$), $3.3\times10^{-4}$ ($200$), $4.0\times10^{-4}$ ($0\bar{2}0$) and $2.9 \times10^{-4}$ ($002$).

\begin{figure} 
    \centering
    \includegraphics[height=7cm]{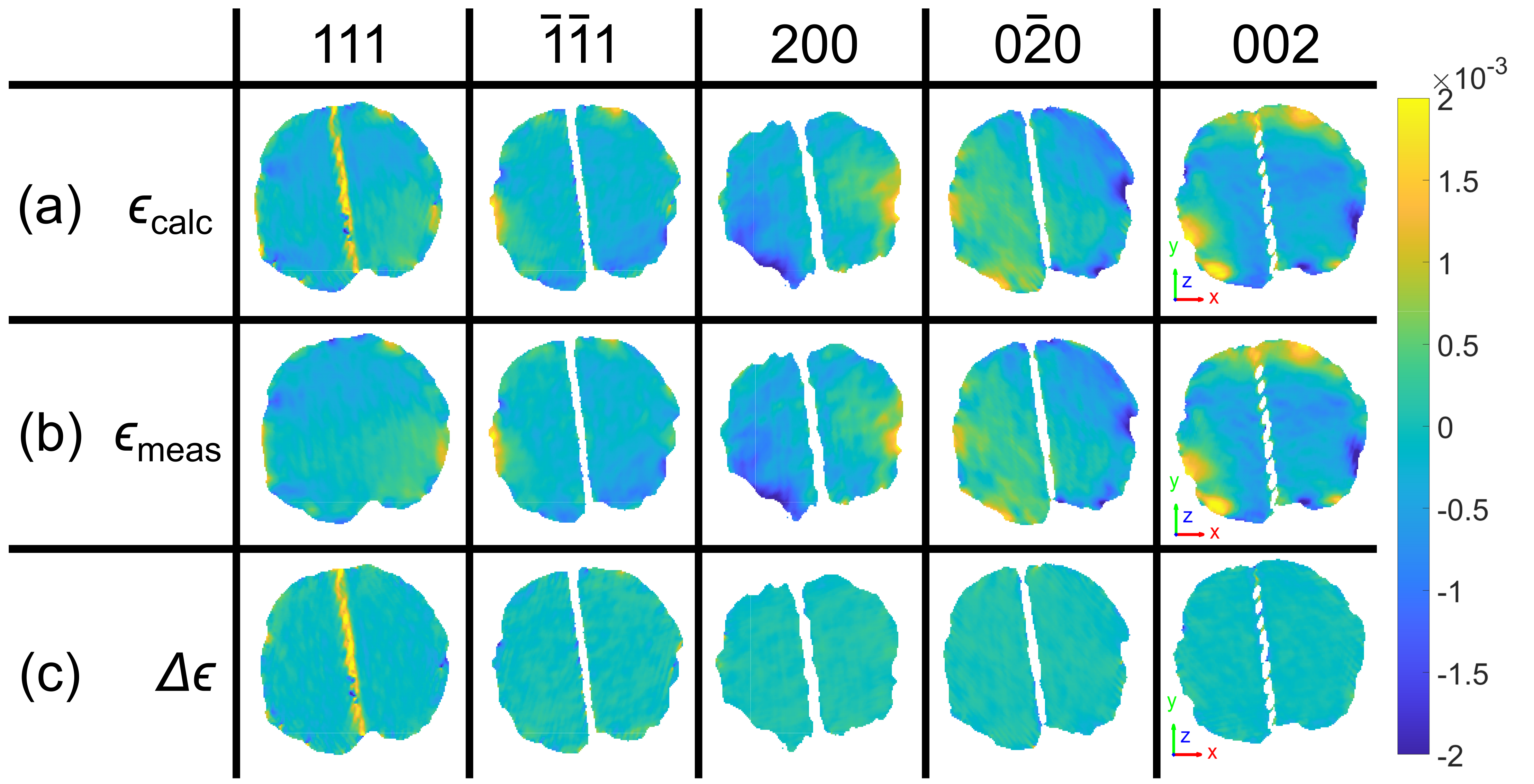}
    \caption{A comparison between the (a) calculated and (b) measured strain for the $111$, $\bar{1}\bar{1}1$, $200$, $0\bar{2}0$, and $002$ reflections at $y = 2.5\ \mathrm{nm}$. The calculated strain is computed using the strain tensor determined using all five reflections following the methodology presented in article. (c) shows the measured strain subtracted from the calculated strain. The amplitude threshold for the reconstructions is 0.30 and the size of the coordinate axes is 100 nm.}
    \label{fig:Strain_error_all}
\end{figure}

If all five reflections are used to predict the strain along the scattering vector for each reflection (Fig. \ref{fig:Strain_error_all}), the average strain error for each reflection, listed in parentheses, is $2.0\times10^{-4}$ ($111$), $1.2\times10^{-4}$ ($\bar{1}\bar{1}1$), $8.2\times10^{-5}$ ($200$), $1.0\times10^{-4}$ ($0\bar{2}0$) and $9.6\times10^{-5}$ ($002$).

The average strain error is close to the strain resolution limit for BCDI, $\sim2\times10^{-4}$ \cite{Yang2021, Hofmann2020}, demonstrating that the reconstructed tensors using all five reflections is accurate. The use of all five reflections reduces the average strain error by 62\%, compared with the use of four reflections.

\section{EDX results}
\label{appendix:EDX_results}
\begin{figure} 
    \centering
    \includegraphics[height=4cm]{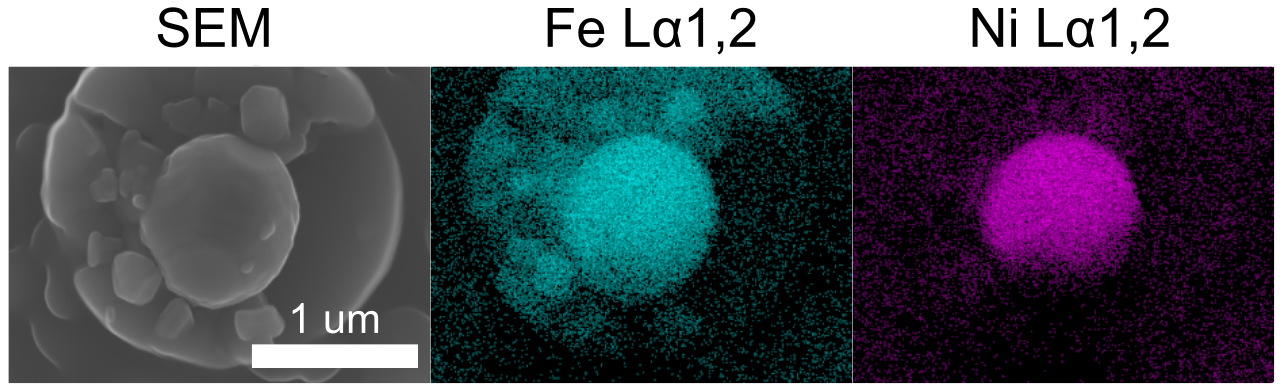}
    \caption{EDX elemental analysis maps of a Fe-Ni crystal. Shown are 2D signals for the primary Fe-Ni emission lines. 
    }
    \label{fig:EDX_map}
\end{figure}

Fig. \ref{fig:EDX_map} shows homogeneous elemental distribution of crystal 2B corresponding to the SEM spectrum in Fig. \ref{fig:SEM_pic}(c).

\section{Supplementary video descriptions} \label{appendix:supplementary_videos}
A brief description of each supplementary video is listed below.
\begin{itemize}
\item SV1 shows x-y plane slices through the lattice strain and rotation tensors in Fig. \ref{fig:Tensor}.
\item SV2 shows y-z plane slices through the lattice strain and rotation tensors in Fig. \ref{fig:Tensor}.
\item SV3 shows z-x plane slices through the lattice strain and rotation tensors in Fig. \ref{fig:Tensor}.
\end{itemize}
\
\


\ack{\textbf{Acknowledgements}} \label{acknowledgements}
\sloppy D.Y, G.H., N.W.P, and F.H. acknowledge funding from the European Research Council under the European Union's Horizon 2020 research and innovation programme (grant agreement No 714697 and N.W.P under the Marie Skłodowska-Curie Actions grant agreement No. 884104 PSI-FELLOW-III-3i). K.S. acknowledges funding from the General Sir John Monash Foundation. The authors acknowledge use of characterisation facilities at the David Cockayne Centre for Electron Microscopy, Department of Materials, University of Oxford and use of the Advanced Research Computing (ARC) facility at the University of Oxford \cite{Richards2015}. X-ray diffraction experiments were performed at the Advanced Photon Source, a US Department of Energy (DOE) Office of Science User Facility operated for the DOE Office of Science by Argonne National Laboratory under Contract No. DE-AC02-06CH11357.
\pagebreak

\bibliographystyle{iucr}
\bibliography{iucr}
\pagebreak





\end{document}